%% file: 0_Main.tex
\documentclass[12pt]{iopart}

\usepackage{bm}
\usepackage[colorlinks=true,allcolors=blue]{hyperref}
\usepackage{graphicx}
\usepackage{subcaption}
\usepackage{float} 
\usepackage{setspace}

\begin{document}
\eqnobysec
\title[]{Effect of triangularity on plasma turbulence and the SOL-width scaling in L-mode diverted tokamak configurations}

\author{K. Lim$^1$, M. Giacomin$^2$, P. Ricci$^1$, A. Coelho$^1$, O. F\'evrier$^1$, D. Mancini$^1$,  D. Silvagni$^3$, and L. Stenger$^1$}

\address{$^1$ Ecole Polytechnique Fédérale de Lausanne (EPFL), Swiss Plasma Center, EPFL SB SPC, Station 13, CH-1015 Lausanne, Switzerland}
\address{$^2$ York Plasma Institute, University of York, York YO10 5DD, United Kingdom}
\address{$^3$ Max-Planck-Institut für Plasmaphysik, Boltzmannstr. 2, D-85748 Garching, Germany}
\ead{kyungtak.lim@epfl.ch}
\vspace{10pt}

\begin{abstract}
    The effect of triangularity on tokamak boundary plasma turbulence is investigated by using global, flux-driven, three-dimensional, two-fluid simulations. The simulations show that negative triangularity stabilizes boundary plasma turbulence, and linear investigations reveal that this is due to a reduction of the magnetic curvature drive of interchange instabilities, such as the resistive ballooning mode. As a consequence, the pressure decay length $L_p$, related to the SOL power fall-off length $\lambda_q$, is found to be affected by triangularity. Leveraging considerations on the effect of triangularity on the linear growth rate and nonlinear evolution of the resistive ballooning mode, the analytical theory-based scaling law for $L_p$ in L-mode plasmas, derived by Giacomin \textit{et al.} [{Nucl. Fusion}, \href{https://doi.org/10.1088/1741-4326/abf8f6}{\textbf{61} 076002} (2021)], is extended to include the effect of triangularity. The scaling is in agreement with nonlinear simulations and a multi-machine experimental database, which include recent TCV discharges dedicated to the study of the effect of triangularity in L-mode diverted discharges. Overall, the present results highlight that negative triangularity narrows the $L_p$ and considering the effect of triangularity is important for a reliable extrapolation of $\lambda_q$ from present experiments to larger devices.
\end{abstract}

%
%
%
%
%
\newpage
\input{1_Intro}
\input{2_GBS_model}

\input{3_NL_analysis}
\input{4_Lp_estimate}
\input{5_Lp_comparison}
\input{6_Conclusion}

\input{7_Acknowledgements}

\appendix
\input{8_Appendix_1}
\input{8_Appendix_2}
\newpage
\bibliographystyle{unsrt}
\bibliography{9_bib.bib}
\end{document}

%% file: 1_Intro.tex
\section{Introduction}
The shape of the plasma cross-section plays an important role in determining the performance of a tokamak. An elongated D-shape plasma was introduced in JET 
based on its improved MHD stability properties \cite{JET, Lazarus1991} and enhanced confinement time obtained by operating in H-mode conditions \cite{Miura1992}. In H-mode plasmas, however, the formation of a steep pressure gradient in the edge region, also known as \textit{pedestal}, often yields transient edge-localized modes (ELMs) that release a large amount of energy ($\sim$ MJ) across the separatrix \cite{Loarte2007}. ELMs can severely degrade the performance of the tokamak as they lead to detrimental impurity accumulation in the core caused by the erosion of the wall materials \cite{Ran1989}.

Recently, the negative triangularity (NT) scenario has attracted increasing attention as an alternative to H-mode operation in positive triangularity (PT) \cite{Kikuchi2019}. The first detailed study focused on the dependence of the confinement time on triangularity, $\delta$, was carried out in TCV with auxiliary electron cyclotron heating in L-mode operation \cite{Weisen1997, Pochelon1999}. The scaling law for energy confinement, found to obey $\tau_E \propto (1+\delta)^{-0.35}$, indicates the increase of energy confinement time in NT plasma, $\delta <0$, with respect to PT scenario, $\delta >0$. Following this initial investigation, a large number of experimental studies on NT were carried out on the TCV \cite{Camenen2007, Fontana2018, Huang2019, Fontana2020, Coda2022}, DIII-D \cite{Austin2019, Marinoni2019, Marinoni2021_2} and ASDEX Upgrade (AUG) \cite{Happel2023} tokamaks, showing H-mode like confinement ($H_{98y2}=1.3$) and ITER-relevant beta ($\beta_N=2.7$) in an intrinsically ELM-free L-mode regime. The effects of plasma shaping on confinement time were actively investigated also through first-principle numerical codes. Both gyrokinetic and gyrofluid simulations \cite{Marinoni2009, Wan2015, Merlo2019, Merlo2021, Duff2022} shed light on the stabilizing effect of negative triangularity over both the ion temperature gradient (ITG) mode and the trapped electron mode (TEM). 

Turning to the boundary, recent works with fluid models \cite{Riva2017, Laribi2021} provide insight on edge plasma turbulence in NT plasma, showing a reduction of the power fall-off length $\lambda_q$ for $\delta<0$. The observed reduction in the SOL width \cite{Faitsch2018} is interpreted as the result of the suppression of plasma turbulence, but recent experimental work also shows that the plasma interactions with the first wall might contribute to a narrower $\lambda_q$ in NT than in PT configurations \cite{Han2021}. As a matter of fact, the reduced $\lambda_q$ in the SOL, which might be a drawback for the use of NT configurations in future fusion reactors, calls for a careful analysis of plasma turbulence in NT scenarios.

The purpose of this paper is to explore the effect of triangularity on edge plasma turbulence and, as a consequence, on the scaling law of the power fall-off length $\lambda_q$. The present work leverages previous simulations \cite{Riva2017, Riva2020}, carried out with the GBS code, that reveal the stabilizing effect of NT on SOL plasma turbulence in a limited configuration. In the present paper, we extend these investigations to consider a diverted configuration, taking into account the interplay existing between the core, edge and SOL regions. We first discuss the results of global, flux-driven, nonlinear, three-dimensional, two-fluid GBS simulations in PT and NT magnetic geometries, particularly in view of the stabilizing effects of NT on edge plasma turbulence. Second, we analyse the effect of plasma shaping on turbulence, deriving a theoretical scaling law for the pressure gradient length $L_p$ that extends the work presented in Refs. \cite{Giacomin2020, Giacomin2021_2} to include the effect of plasma triangularity. Finally, the derived scaling law is compared with nonlinear simulations and a multi-machine experimental database that includes recent discharges carried out on the TCV tokamak to study the effect of triangularity on plasma turbulence, as well as discharges from the ASDEX Upgrade \cite{Silvagni2020}, Alcator C-Mod \cite{Brunner2018}, COMPASS \cite{Horacek2020}, JET \cite{Scarabosio2013} and MAST \cite{Ahn2006} tokamaks for L-mode plasmas. 

We focus on the sheath-limited regime, characterized by a small temperature gradient between upstream (i.e., outboard midplane) and the divertor targets, in contrast to the conduction-limited regime \cite{Stangeby2000}. The two regimes can be identified by using the SOL collisionality parameter $\nu^*=10^{-16}n_u L /T_u^2$ derived from the two-point model \cite{Stangeby2000}, where $u$ denotes the upstream quantities and $L$ is the connection length. In general, the sheath-limited regime is characterized by weak collisionality ($\nu^*<10$) while a significant temperature drop is often observed in the conduction-limited regime ($\nu^*>15$).

The remainder of this paper is organized as follows. In Section \ref{Sec2}, we introduce the physical model to study boundary plasma turbulence. Then, the results of nonlinear GBS simulations are discussed in Section \ref{Sec3} focusing on the stabilizing effect of NT on edge plasma turbulence. The theoretical derivation of the scaling law of the pressure decay length, $L_p$, which takes into account plasma shaping parameters, is detailed in Section \ref{Sec4} and comparisons with nonlinear simulations results are presented. In Section \ref{Sec5}, the validity of our newly derived scaling law is tested against a multi-machine experimental database. Finally, the conclusions are drawn in Section \ref{Sec6}.

%% file: 2_GBS_model.tex
\section{Numerical model}\label{Sec2}
The high plasma collisionality ($L_\parallel \gg \lambda_e$, $L_\parallel$ being the parallel length scales of turbulent modes and $\lambda_e$ the electron mean-free path) justifies the use of the two-fluid Braginskii model to study boundary plasma turbulence in L-mode discharges. In addition, the drift limit of the Braginskii model \cite{Zeiler1997} can be considered since plasma turbulence in the boundary region occurs on time scales slower than $1/\Omega_{ci}$, being $\Omega_{ci}=eB/m_i$ the ion cyclotron frequency. 

Initially developed to study turbulence in basic plasma physics experiments \cite{Rogers2013} and limited tokamak configurations, the GBS code \cite{Ricci2012, Halpern2016, Giacomin2022} solves the drift-reduced Braginskii equations to evolve plasma turbulence in the tokamak boundary. The implementation of a spatial discretization algorithm independent of the magnetic field \cite{Paruta2018} allows GBS to simulate diverted configurations with an arbitrary magnetic equilibrium \cite{Beadle2020, Giacomin2020_2}, as well as non-axisymmetric configurations, such as the stellarators \cite{Coelho2022}. While the plasma model implemented in GBS was developed in recent years to include the neutral dynamics \cite{Wersal2015}, we do not include it in the simulations presented here, therefore focusing on the sheath-limited regime. We also neglect electromagnetic effects that can be important at high values of plasma beta \cite{Giacomin2022_2}. Accordingly, the GBS equations considered in the present study can be written in dimensionless form as:
\begin{equation}
    \fl \frac{\partial n}{\partial t} = -\frac{\rho_*^{-1}}{B}[\phi, n] + \frac{2}{B}\bigg[C(p_e)-nC(\phi)\bigg]-\nabla_\parallel (nv_{\parallel e}) + D_n \nabla^2_\perp n + s_n,
\label{GBS_density}
\end{equation}
\begin{equation}
    \fl\frac{\partial \Omega}{\partial t} = -\frac{\rho_*^{-1}}{B} \nabla \cdot [\phi, \omega] -\nabla \cdot \big(v_{\parallel i} \nabla_\parallel\omega\big) + B^2 \nabla_\parallel j_\parallel + 2B C(p_e + \tau p_i) + \frac{B}{3}C(G_i) + D_\Omega \nabla_\perp^2 \Omega,
\label{GBS_vorticity}
\end{equation}
\begin{equation}
    \fl\frac{\partial v_{\parallel i}}{\partial t} = -\frac{\rho_*^{-1}}{B}[\phi, v_{\parallel i}]-v_{\parallel i} \nabla_\parallel v_{\parallel i} - \frac{1}{n}\nabla_\parallel (p_e + \tau p_i) -\frac{2}{3n}\nabla_\parallel G_i + D_{v_{\parallel i}}\nabla_\perp^2 v_{\parallel i},
\end{equation}
\begin{eqnarray}
    \fl \frac{\partial v_{\parallel e}}{\partial t} &= -\frac{\rho_*^{-1}}{B}[\phi, v_{\parallel, e}] - v_{\parallel e}\nabla_\parallel v_{\parallel e} + \frac{m_i}{m_e}\Bigg(\nu j_\parallel + \nabla_\parallel \phi -\frac{1}{n}\nabla_\parallel p_e-0.71 \nabla_\parallel T_e -\frac{2}{3n}\nabla_\parallel G_e \Bigg) \nonumber \\
    \fl& + D_{v_{\parallel e}}\nabla_\perp^2 v_{\parallel e}, 
\end{eqnarray}
\begin{eqnarray}
    \fl \frac{\partial T_i}{\partial t} &=-\frac{\rho_*^{-1}}{B}[\phi, T_i] - v_{\parallel i}\nabla_\parallel T_i + \frac{4}{3}\frac{T_i}{B}\Bigg[C(T_e) + \frac{T_e}{n}C(n)-C(\phi) \Bigg] -\frac{10}{3}\tau \frac{T_i}{B}C(T_i) \nonumber \\
    \fl &+ \frac{2}{3}T_i\Bigg[(v_{\parallel i}-v_{\parallel e})\frac{\nabla_\parallel n}{n}-T_i \nabla_\parallel v_{\parallel e}\Bigg] + 2.61\nu n (T_e -\tau T_i)+ \nabla_\parallel( \chi_{\parallel i}\nabla_\parallel T_i) \nonumber\\
    \fl &+ D_{T_i}\nabla_\perp^2 T_i + s_{T_i},
\end{eqnarray}
\begin{eqnarray}
    \fl \frac{\partial T_e}{\partial t} &= -\frac{\rho_*^{-1}}{B}[\phi, T_e] - v_{\parallel e}\nabla_\parallel T_e + \frac{2}{3}T_e\Bigg[0.71 \frac{\nabla_\parallel j_\parallel}{n}-\nabla_\parallel v_{\parallel e}\Bigg] -2.61\nu n (T_e-\tau T_i)
    \nonumber \\
    \fl&+ \frac{4}{3}\frac{T_e}{B}\Bigg[\frac{7}{2}C(T_e)+\frac{T_e}{n}C(n)-C(\phi) \Bigg] + \nabla_\parallel (\chi_{\parallel e}\nabla_\parallel T_e) + D_{T_e}\nabla_\perp^2 T_e + s_{T_e}. \label{GBS_electron_temperature}
\end{eqnarray}
Equations (\ref{GBS_density}-\ref{GBS_electron_temperature}) are closed by the evaluation of the electrostatic potential that avoids the Boussinesq approximation,
\begin{eqnarray}
    \fl \nabla \cdot \bigg( n\nabla_\perp \phi \bigg) = \Omega - \tau \nabla_\perp^2 p_i,
\label{GBS_Poisson}
\end{eqnarray}
where $\Omega=\nabla \cdot \omega = \nabla \cdot (n\nabla_\perp \phi + \tau \nabla_\perp p_i)$ is the scalar vorticity.

In Eqs. (\ref{GBS_density}-\ref{GBS_electron_temperature}) and in the remainder of this paper, the plasma density $n$, the ion and electron temperatures, $T_{i}$ and $T_{e}$, the ion and electron parallel velocities, $v_{\parallel i}$ and $v_{\parallel e}$, and the electrostatic potential $\phi$ are normalized to the reference values $n_0, T_{i0}, T_{e0}, c_{s0}=\sqrt{T_{e0}/m_i}, c_{s0}$, and $T_{e0}/e$, respectively. The perpendicular lengths are normalized to the ion sound Larmor radius, $\rho_{s0}=c_{s0}/\Omega_{ci}$, and parallel lengths are normalized to the tokamak major radius, $R_0$. Time is normalized to $t_0=R/c_{s0}$. In addition, the dimensionless parameters that determine the plasma dynamics in Eqs. (\ref{GBS_density}-\ref{GBS_Poisson}) are the normalized ion sound Larmor radius, $\rho_*=\rho_{s0}/R_0$, the ratio of the ion to the electron temperature, $\tau=T_{i0}/T_{e0}$, the normalized ion and electron viscosities, $\eta_{0,i}$ and $\eta_{0,e}$, the normalized ion and electron parallel thermal conductivities $\chi_{\parallel i}$ and $\chi_{\parallel e}$, and the normalized Spitzer resistivity $\nu=e^2n_0R_0/(m_ic_{s0}\sigma_\parallel)=\nu_0 T_e^{-3/2}$, with
\begin{equation}
    \sigma_\parallel = \Bigg(1.96\frac{n_0 e^2 \tau_e}{m_e}\Bigg)n = \Bigg[\frac{5.88}{4\sqrt{2\pi}} \frac{(4\pi \epsilon_0)^2}{e^2}\frac{T_{e0}^{3/2}}{\lambda \sqrt{m_e}}\Bigg]T_e^{3/2}
\end{equation}
and, as a consequence,
\begin{equation}
    \nu_0 =\frac{4\sqrt{2\pi}}{5.88} \frac{e^4}{(4\pi \epsilon_0)^2}\frac{\sqrt{m_e}R_0 n_0 \lambda}{m_i c_{s0}T_{e0}^{3/2}},
\label{Collisionality}
\end{equation}
where $\lambda$ is the Coulomb logarithm.
The gyroviscous terms are defined as 
\begin{eqnarray}
    G_i=-\eta_{0i}\Bigg[2\nabla_\parallel v_{\parallel i} + \frac{1}{B}C(\phi) + \frac{1}{enB}C(p_i) \Bigg]
\end{eqnarray}
and
\begin{eqnarray}
    G_e=-\eta_{0e}\Bigg[2\nabla_\parallel v_{\parallel e} + \frac{1}{B}C(\phi)-\frac{1}{enB}C(p_e) \Bigg],
\end{eqnarray}
where $\eta_{0i}=0.96nT_i\tau_i$ and $\eta_{0e}=0.73nT_e\tau_e$. The diffusion terms $D_f\nabla_\perp^2 f$ are added on the right hand side of Eqs. (\ref{GBS_density}-\ref{GBS_electron_temperature}) to improve the numerical stability of the simulations.

The spatial operators that appear in Eqs. (\ref{GBS_density}-\ref{GBS_Poisson}) are the Poisson bracket operator, $[f,g]=\bm{b}\cdot(\nabla g \times \nabla f)$, the curvature operator, $\mathcal{C}(f)=B[\nabla \times (\bm{b}/B)]/2 \cdot \nabla f$, the parallel gradient operator, $\nabla_\parallel f = \bm{b} \cdot \nabla f$, and the perpendicular Laplacian operator, $\nabla_\perp^2 f = \nabla \cdot [(\bm{b}\times \nabla f)\times \bm{b}]$, where $\bm{b}=\bm{B}/B$ is the unit vector of the magnetic field. It is useful to represent these operators in tensorial form for an arbitrary magnetic field \cite{Dhaeseleer1991},
\begin{eqnarray}
    [\phi, f] &=\frac{1}{\mathcal{J}}\epsilon_{ijk}b_i \frac{\partial \phi}{\partial \xi^j}\frac{\partial f}{\partial \xi^k}, \label{Operators_Poisson} \\
    \nabla_\parallel f &=b^j \frac{\partial f}{\partial \xi^j},\\
    \mathcal{C}(f) &=\frac{B}{2\mathcal{J}}\frac{\partial c_m}{\partial \xi^j}\frac{\partial f}{\xi^k}\epsilon_{kjm}, \\
    \nabla_\perp^2f &=\frac{1}{\mathcal{J}}\frac{\partial}{\partial \xi^k}\bigg( \mathcal{J}^{-1}\epsilon_{klm}\epsilon_{i\alpha\beta}g_{mi}b_lb_\alpha \frac{\partial f}{\partial \xi^\beta}\bigg), \label{Operators_divergence}
\end{eqnarray}
where the Einstein convention is used with the Levi-Civita symbol $\epsilon_{ijk}$, and we introduce an arbitrary set of coordinates $\bm{\xi}=(\xi^1, \xi^2, \xi^3)$, the coefficients $c_m=b_m/B$, and $b_i=g_{ij}b^j$, the covariant metric tensor $g^{ij}=\nabla \xi^i \cdot \nabla \xi^j$ and the Jacobian $\mathcal{J}=1/\sqrt{\textrm{det}(g^{ij})}$. In addition, we express $\nabla \cdot \bm{b} =\frac{1}{\mathcal{J}}\frac{\partial}{\partial \xi^i}(b^i \mathcal{J})$.  

In GBS, the axisymmetric magnetic field is represented as $\bm{B}=RB_\varphi\nabla \varphi + \nabla \varphi \times \nabla \psi$ where the poloidal flux function $\psi$ and the toroidal angle $\varphi$ are introduced. The non-field-aligned cylindrical coordinates $(R, \varphi, Z)$ is used, where $R$ corresponds to the radial distance from the tokamak symmetry axis and $Z$ is the vertical coordinate. By considering the large aspect ratio limit ($\epsilon \ll 1$), assuming $\delta \sim B_p/B_\varphi \ll 1$, and retaining only the leading order terms in $\epsilon$ and $\delta$, the differential operators in Eqs. (\ref{Operators_Poisson}-\ref{Operators_divergence}) implemented in GBS can be recast as  
\begin{eqnarray}
    [\phi, f] &=\frac{B_\varphi}{B}(\partial_Z \phi \partial_R f - \partial_R \phi \partial_Z f), \\
    \nabla_\parallel f &=\partial_Z \psi \partial_R f - \partial_R\psi \partial_Z f + \frac{B_\varphi}{B}\partial_\varphi f,\\
    \mathcal{C}(f) &=\frac{B_\varphi}{B}\partial_Z f, \\
    \nabla_\perp^2f &=\partial^2_{RR}f + \partial^2_{ZZ}f.
\end{eqnarray}
The presence of the density source in the proximity of the last closed flux surface (LCFS) mimics the ionization of neutral atoms and the temperature source the Ohmic heating in the core. The analytical expressions of source terms are expressed as
\begin{equation}
    s_n = s_{n0}\exp{\Bigg\{ -\frac{[\psi(R,Z)-\psi_n]^2}{\Delta_n^2}\Bigg\}}
\end{equation}
and
\begin{equation}
    s_T = \frac{s_{T0}}{2}\Bigg[\tanh\Bigg[-\frac{\psi(R,Z)-\psi_T}{\Delta_T}\Bigg]+1\Bigg],
\end{equation}
where $\psi_n$ and $\psi_T$ represent two flux surfaces located inside the LCFS, while $\Delta_n$ and $\Delta_T$ determine the radial width of the source terms.

The boundary conditions imposed at the magnetic pre-sheath, where the ion drift approximation is not valid, are derived in Ref. \cite{Loizu2012} to generalize the Bohm-Chodura criterion, and are adapted to the diverted configuration \cite{Paruta2018}. By neglecting terms associated with the plasma gradients along the wall, these boundary conditions for the top and bottom walls can be expressed as 
\begin{equation}
    v_{\parallel i}=\pm \sqrt{T_e+\tau T_i},
\end{equation}
\begin{equation}
    v_{\parallel e}=\pm\sqrt{T_e + \tau T_i}\,\mathrm{max}\Bigg[\exp{\Bigg( \Lambda - \frac{\phi}{T_e}\Bigg), \exp{(\Lambda)}} \Bigg],
\end{equation}
\begin{equation}
    \partial_Zn = \mp \frac{n}{\sqrt{T_e+\tau T_i}}\partial_Z v_{\parallel i},
\end{equation}
\begin{equation}
    \partial_Z \phi = \mp \frac{T_e}{\sqrt{T_e + \tau T_i}}\partial_Z v_{\parallel i},
\end{equation}
\begin{equation}
    \partial_Z T_e = \partial_Z T_i = 0,
\end{equation}
\begin{equation}
    \omega = -\frac{T_e}{T_e + \tau T_i}\bigg[(\partial_Z v_{\parallel i})^2 \pm \sqrt{T_e + \tau T_i} \partial^2_Z v_{\parallel i}\bigg],
\end{equation}
where the $\pm$ sign indicates the magnetic field lines entering (top sign) or leaving (bottom sign) the wall and $\Lambda\simeq 3$. Moreover, the electric potential is chosen to be $\phi=\Lambda T_e$ at the left and right walls of the simulation domain, and vanishing perpendicular derivatives to the wall are set for the other quantities. 

In order to analyse the simulation results, we make use of a linear solver based on a local flux-tube coordinate system. This system of coordinates is based on the toric coordinate system $(\Psi=r, \Theta=a\theta_*, \zeta=R_0\varphi)$ where $a$ is the minor radius and the straight-field-line angle $\theta_*$ is defined as
\begin{equation}
    \theta_* = \frac{1}{q(r)}\int_0^\theta \frac{\bm{B}\cdot \nabla \varphi}{\bm{B}\cdot \nabla \theta'} d\theta',
\label{field_line_angle}
\end{equation}
being $\theta$ and $\theta'$ the poloidal angle, and $q(r)$ the safety factor
\begin{equation}
    q(r) = \frac{1}{2\pi}\int_0^{2\pi}\frac{\bm{B}\cdot \nabla \varphi}{\bm{B} \cdot \nabla \theta}d\theta.
\end{equation}
This field-aligned system is then transformed into the flux-tube coordinates $(r, \alpha, \theta_*)$ where $\alpha=\varphi-q(r)\theta_*$ is a field line label, and finally rescaled into the local flux-tube coordinates as $x=r, y=(a/q)\alpha, z=q R_0 \theta_*$. The $(x,y)$ plane is perpendicular to the magnetic field, and $z$ is a field-aligned coordinate. In the rescaled flux-tube coordinate system, the geometrical operators in Eqs. (\ref{Operators_Poisson}-\ref{Operators_divergence}) can be rewritten as
\begin{eqnarray}
    [\phi,f]&=\mathcal{P}_{xy}[\phi, f]_{xy} + \mathcal{P}_{yz}[\phi,f]_{yz}+\mathcal{P}_{zx}[\phi,f]_{zx}, \label{Operator_Poisson}\\
    \nabla_\parallel f &= \mathcal{D}^x \frac{\partial f}{\partial x} + \mathcal{D}^y\frac{\partial f}{\partial y} + \mathcal{D}^z \frac{\partial f}{\partial z}, \\
    \mathcal{C}(f) &= C^x \frac{\partial f}{\partial x} + C^y \frac{\partial f}{\partial y} + \mathcal{C}^z \frac{\partial f}{\partial z}, \\
    \nabla_\perp^2 f &= \mathcal{N}^x \frac{\partial f}{\partial x} + \mathcal{N}^y \frac{\partial f}{\partial y} + \mathcal{N}^z \frac{\partial f}{\partial z} + \mathcal{N}^{xx}\frac{\partial^2 f}{\partial x^2} + \mathcal{N}^{xy}\frac{\partial^2 f}{\partial x \partial y} \label{Operator_Curvature}  \\
    & + \mathcal{N}^{yy}\frac{\partial^2f}{\partial y^2} + \mathcal{N}^{xz}\frac{\partial^2 f}{\partial x \partial z} + \mathcal{N}^{yz}\frac{\partial^2 f}{\partial y \partial z} + \mathcal{N}^{zz}\frac{\partial^f}{\partial z^2}, \label{Operator_Laplacian}
\end{eqnarray}
where the coefficients appearing in front of the spatial derivatives are computed as a function of plasma shaping parameters. Detailed expressions of these coefficients are derived in \ref{Appendix:coef}.

%% file: 3_NL_analysis.tex
\section{Nonlinear analysis}\label{Sec3}
We describe a set of nonlinear GBS simulations used to investigate the effect of triangularity on boundary plasma turbulence. All simulations presented herein use the numerical grid $N_R \times N_Z \times N_\varphi = 240 \times 320 \times 80$ and a time step $\Delta t=10^{-5} \times R_0/c_{s0}$. The size of the simulation domain is set to $L_R=600 \rho_{s0}, L_Z=800 \rho_{s0}$ and $ \rho_*=\rho_{s0}/R_0=1/700$. As a reference, we note that, considering $B_T=0.9\textrm{T}$ and $T_{e0}=20\textrm{eV}$, the normalizing parameters are $c_{s0}=3.8 \times 10^4\textrm{m/s}$, $\rho_{s0}=0.5 \textrm{mm}$. They yield $L_R \simeq 30\,\textrm{cm}, L_Z \simeq 40 \,\textrm{cm}$ and $R_0 \simeq 25\,\textrm{cm}$, corresponding to a tokamak with 1/3 of the TCV size \cite{Reimerdes2022}, approximately. Other plasma parameters are kept constant throughout the present study, such as $\tau=T_{i0}/T_{e0}=1$, $m_i/m_e=200$, $\eta_{0e}=\eta_{0i}=1$ and $ \chi_{\parallel e}=\chi_{\parallel i}=1$. The direction of the toroidal magnetic field $B_T$ is set for the ion-$\nabla B$ drift being away from the X-point (\textit{unfavorable} direction for H-mode access). In addition, we use $a/R_0=0.3$ and the plasma current on axis is chosen to have a safety factor $q_0 \simeq 1$ at the magnetic axis and $q_{95}\simeq 4$ at the tokamak edge. 

\begin{figure}[H]
\begin{center}
	\subfloat[Negative triangularity ]{\includegraphics[width=0.35\textwidth]{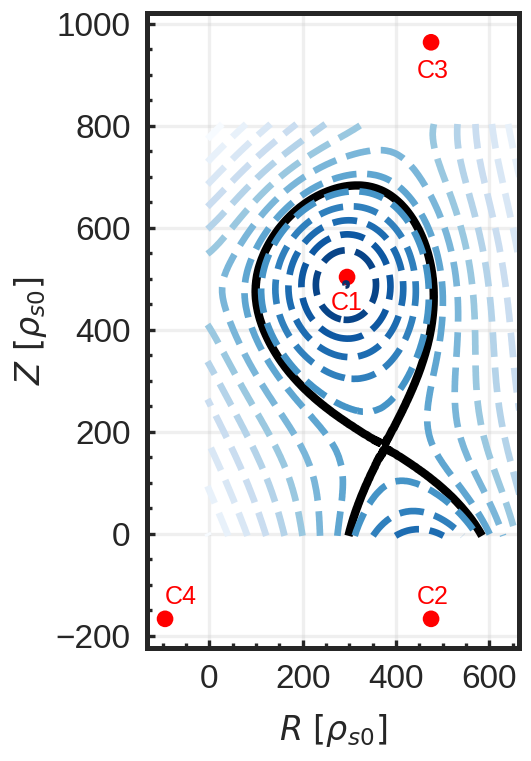}}\label{Neg}
	\subfloat[Positive triangularity]{\includegraphics[width=0.35\textwidth]{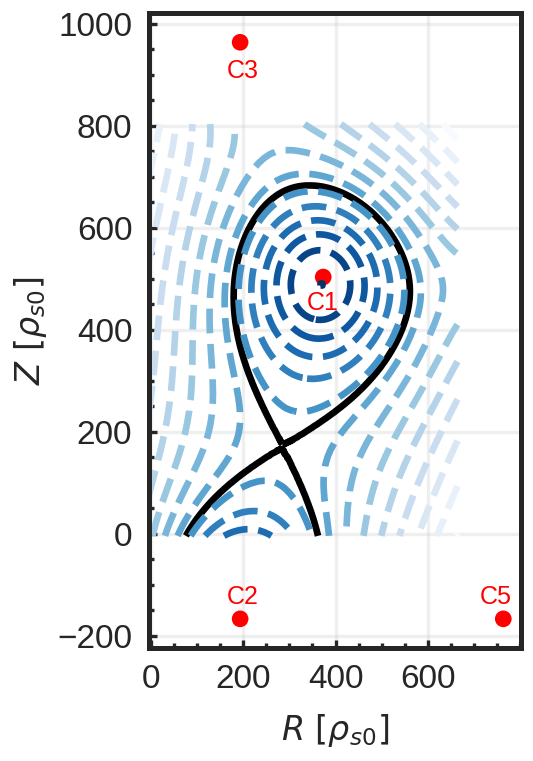}}\label{Pos}
\caption{Magnetic equilibrium profile used for the nonlinear GBS simulations for NT ($\delta\simeq-0.3$) and PT ($\delta \simeq+0.3$) plasmas with $\kappa \simeq 1.3$. Red dots represent the position of the current that generates the magnetic field, i.e. the main plasma current (C1), the divertor current (C2), the upper shaping current (C3), the left shaping current (C4) and the right shaping current (C5).}
\label{Magnetic_equilibrium}
\end{center}
\end{figure}
The magnetic geometries of the NT and PT plasmas that we consider in the present study are shown in Figure \ref{Magnetic_equilibrium}. These equilibria are constructed by solving the Biot-Savart law in the infinite aspect-ratio limit with a Gaussian-like centred current with additional current filaments outside the simulation domain. The magnetic equilibria shown in Figure \ref{Magnetic_equilibrium} are characterized by an elongation $\kappa\simeq1.3$ and triangularity $\delta\simeq\pm 0.3$. Denoting $Z_{max}, R_{max}$ and $Z_{min}, R_{min}$ as the maximum and minimum values of $Z$ and $R$ along the separatrix, the shaping parameters are defined as \cite{Sauter2016} 
\begin{eqnarray}
    \kappa &= \frac{Z_{max}-Z_{min}}{R_{max}-R_{min}}, 
\end{eqnarray}
and
\begin{eqnarray}
    \delta &=\frac{\delta_{upper}+\delta_{lower}}{2},
    \label{shaping_parameters}
\end{eqnarray}
where $\delta_{upper}$ and $\delta_{lower}$ denote the upper and lower triangularity, respectively, being 
\begin{eqnarray}
    \delta_{upper} = \frac{R_0-R(Z=Z_{max})}{a}
\end{eqnarray}
and
\begin{eqnarray}
\delta_{lower} = \frac{R_0 - R(Z=Z_{min})}{a},
\end{eqnarray}
with
\begin{eqnarray}
    R_0 &= \frac{R_{max}+R_{min}}{2}
\end{eqnarray}
and
\begin{eqnarray}
    a = \frac{R_{max} - R_{min}}{2}.
\end{eqnarray}
Recent study carried out on TCV report that decreasing $\delta_{upper}$ reduces the amplitude of turbulent fluctuations and leads to an increased confinement time, while $\delta_{lower}$ affects mostly the plasma turbulence in the boundary near the X-point by modifying the divertor geometry \cite{Faitsch2018}. In the present study, the sign of both $\delta_{lower}$ and $\delta_{upper}$ is reversed when obtaining $\delta=\pm 0.3$.

Different turbulent regimes in the tokamak boundary are observed in GBS, depending on the edge collisionality and input heat power \cite{Giacomin2020}, as they result from different driving instabilities such as the resistive ballooning modes (RBMs) and resistive drift waves (RDWs) \cite{Zeiler1997, Scott1997}. The plasma collisionality and heating source in our simulations are chosen so that our simulations are in the RBM regime, which is equivalent to the L-mode operational regime of tokamaks. In particular, being destabilized mainly by the magnetic field curvature and plasma pressure gradient, RBMs are known to be strongly affected by the plasma shaping, while its impact is less important on RDWs \cite{Riva2017}. 

The simulations described in this paper are analysed in the quasi-steady state regime established when plasma sources, perpendicular transport and losses to the vessel wall balance each other and all quantities fluctuate around constant values. Once the simulations reach this quasi-steady state, all quantities are toroidally averaged over a $10t_0$ time frame to evaluate the equilibrium profiles. In the present paper, fluctuating quantities are expressed with tilde and time- and toroidally-averaged quantities with an overline, e.g. $\phi=\bar{\phi}+\tilde{\phi}$.

\begin{figure}[H]
\begin{center}
	\subfloat[NT ($\delta=-0.3$) plasma]{\includegraphics[width=0.4\textwidth]{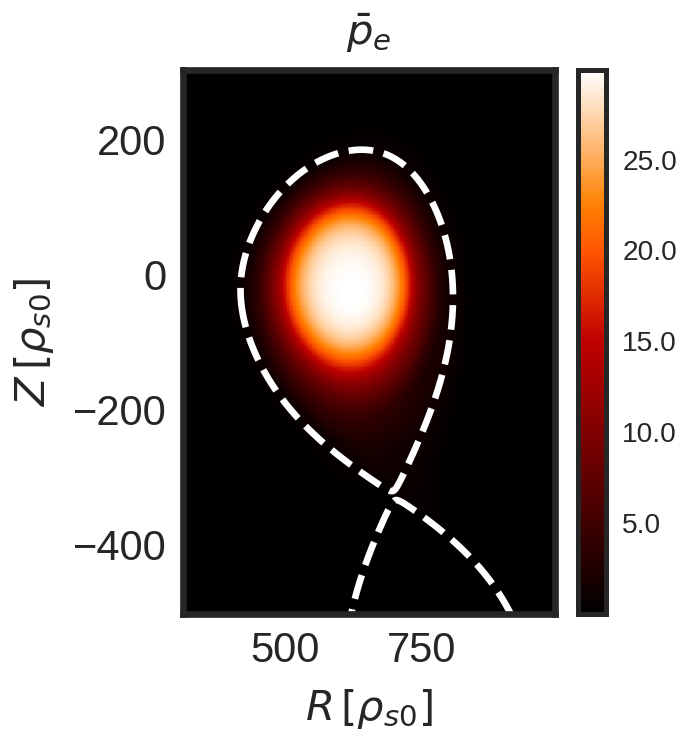}}\label{NT_temp}
	\subfloat[PT ($\delta=+0.3$) plasma]{\includegraphics[width=0.4\textwidth]{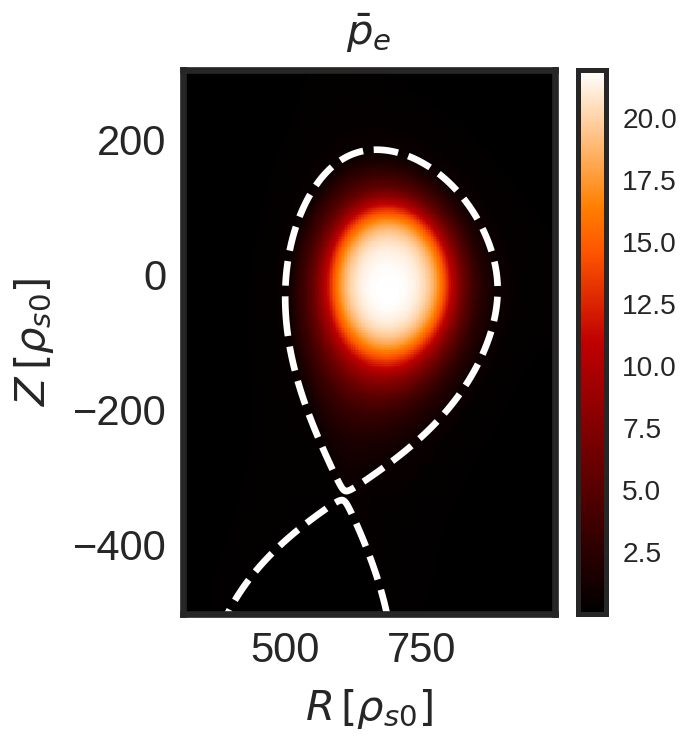}}\label{PT_temp}
\caption{Poloidal cross section of the equilibrium electron pressure for NT and PT plasmas with $s_{T0}=0.025$ and $\nu_0=0.1$. The dashed white line represents the separatrix.}
\label{2D_pressure}
\end{center}
\end{figure}

\begin{figure}[H]
\begin{center}
\includegraphics[width=0.4\textwidth]{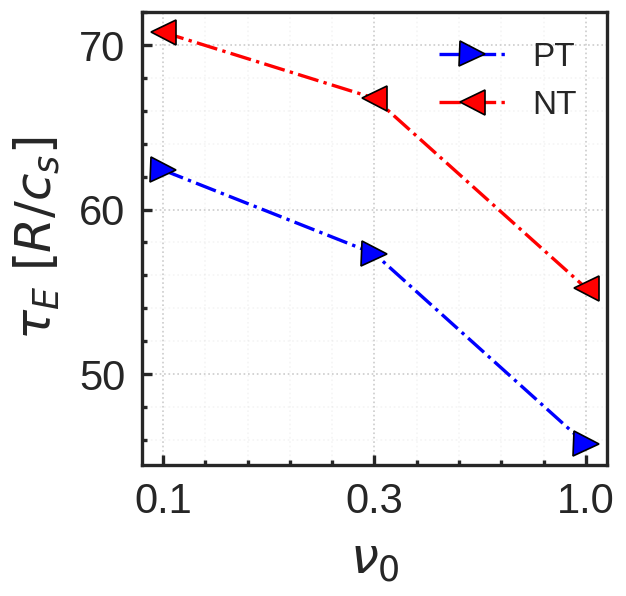}
\caption{Energy confinement time $\tau_E$ as a function of $\nu_0$. The heating source is kept constant ($s_{T0}=0.025$).}
\label{Fig_confinement_time}
\end{center}
\end{figure}

Figure \ref{2D_pressure} displays the equilibrium electron pressure, $\bar{p}_e$, in PT and NT simulations revealing higher plasma pressure in the case of the NT plasma, despite the same sources ($s_{T0}=s_{n0}=0.075$) being used in the two simulations. Higher $\bar{p}_e$ values are associated with a reduced transport level and a higher confinement time. The qualitative estimate of the electron energy confinement time $\tau_E$ is evaluated from the plasma energy content inside LCFS divided by heating power,  
\begin{eqnarray}
    \tau_E = \frac{3}{2}\frac{\int_{A_{LCFS}} \bar{p}_e dRdZ}{\int_{A_{LCFS}} s_p dRdZ},
\end{eqnarray}
and it is shown for different values of edge collisionality in Figure \ref{Fig_confinement_time}. The analysis reveals an improved energy confinement time for $\delta<0$, in agreement with experimental observations from TCV \cite{Camenen2007, Fontana2018, Coda2022} and DIII-D \cite{Austin2019, Marinoni2019}. The energy confinement time for both NT and PT decreases as the collisionality increases. In fact, transport driven by RBMs increases with collisionality leading to reduced energy confinement time for the same value of the input power. 

In agreement with these observations, first-principle simulations based on a gyrokinetic model \cite{Marinoni2019, Merlo2019} show that NT plasmas are characterized by density and temperature fluctuations of reduced amplitude, yielding an enhanced confinement time. However, the reason of the higher confinement is attributed to kinetic effects that stabilize linear instabilities, such as the trapped electron modes and the ion temperature gradient modes. Kinetic effects are not retained in our fluid simulations, limiting their reliability to the study of core phenomena. 

\begin{figure}[H]
\begin{center}
	\subfloat[Density]{\includegraphics[width=0.33\textwidth]{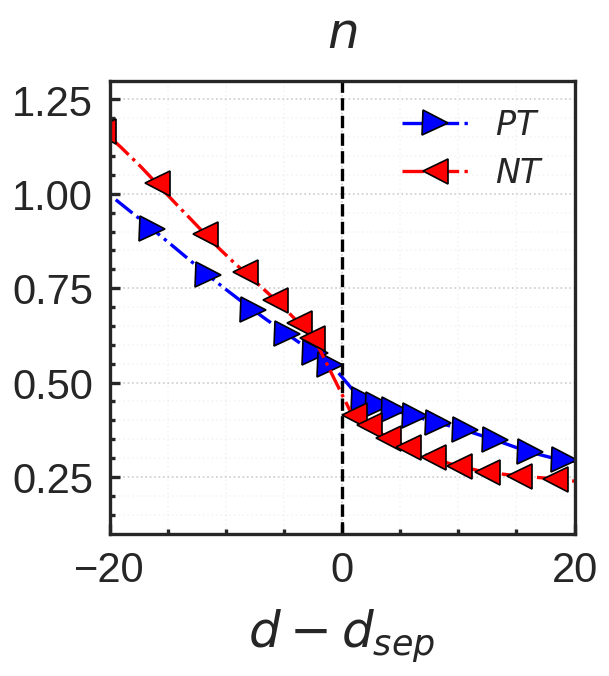}}\label{OMP_Nu01_N}
	\subfloat[Electron temperature]{\includegraphics[width=0.3\textwidth]{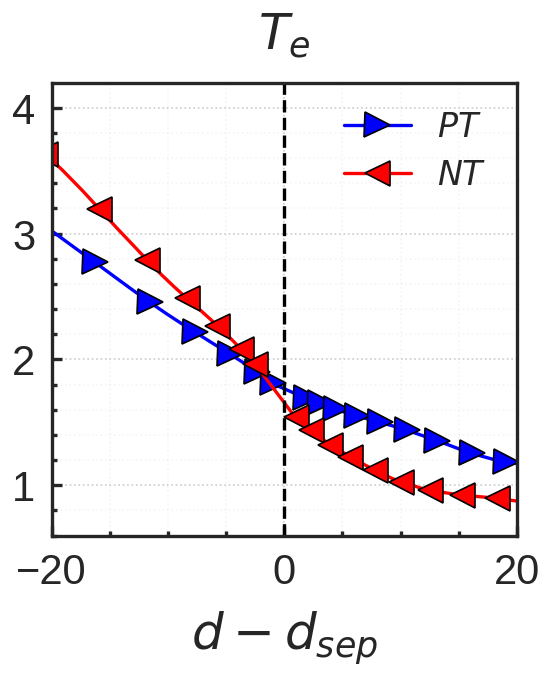}}\label{OMP_Nu01_Te}
	\subfloat[Electron pressure]{\includegraphics[width=0.3\textwidth]{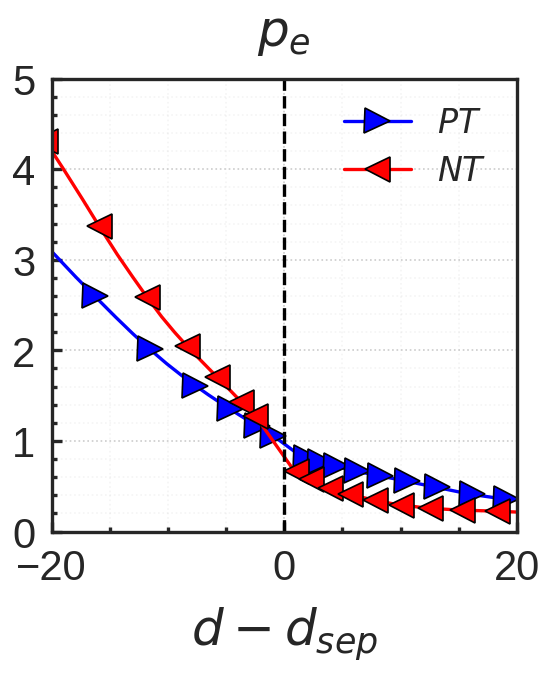}}\label{OMP_Nu01_pe}
\caption{Radial profiles of density (a), electron temperature (b) and electron pressure (c) at the outer midplane for NT and PT plasmas for the simulations with $s_{T0}=0.025$ and $\nu_0=0.1$.}
\label{OMP_profile}
\end{center}
\end{figure}

In Figure \ref{OMP_profile}, typical radial profiles of electron density, temperature and pressure at the outer midplane are presented for a NT and a PT simulation with $s_{T0}=0.025$ and $\nu_0=0.1$. The NT configuration is characterized by steeper equilibrium gradients across the separatrix, particularly evident in the electron temperature and pressure profiles.

\begin{figure}[H]
\begin{center}
	\includegraphics[width=\textwidth]{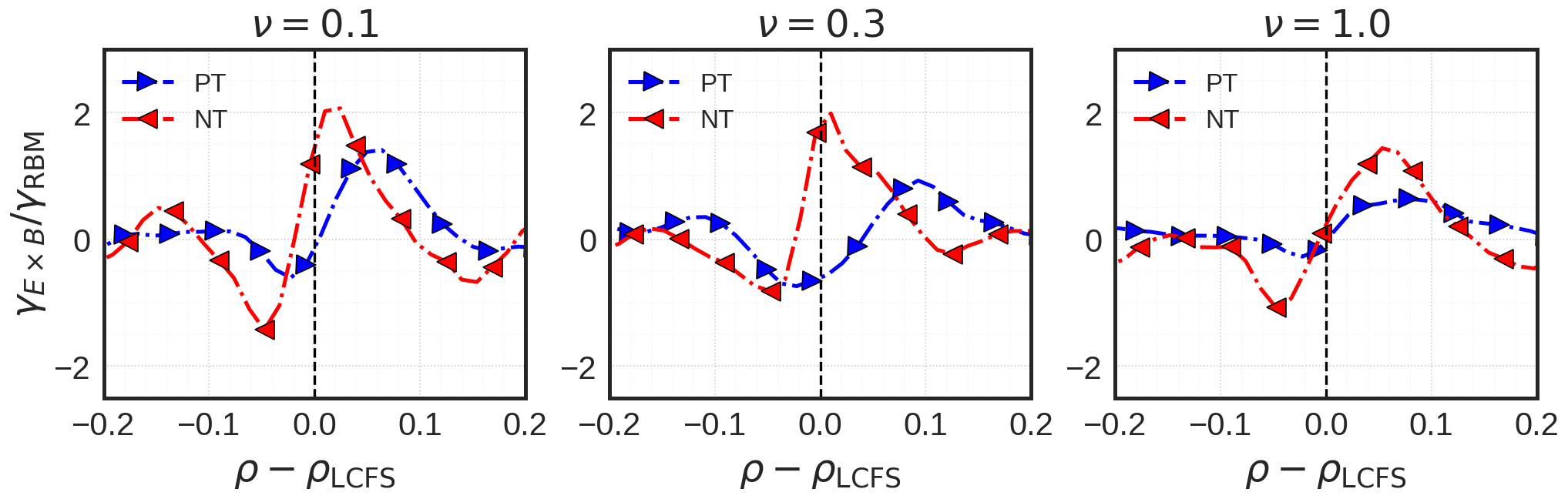}
\caption{$E \times B$ shear rate normalized to $\gamma_{RBM}$, the growth rate of RBMs, for a NT and a PT plasma across the separatrix. The heating source is kept constant ($s_{T0}=0.025$).}
\label{Shear_profile}
\end{center}
\end{figure}

The steeper gradient sustained near the separatrix is associated to a larger $\bm{E} \times \bm{B}$ shear rate, $\gamma_{E \times B} =\rho^{-1}_* \partial^2_r \bar{\phi}$. In Figure \ref{Shear_profile}, radial profiles of the shear rate, normalized to the RBM growth rate, $\gamma_{\rm{RBM}} = \sqrt{(2
\bar{T}_e)/(\rho_* L_p)}$, where $L_p=- p/\nabla p$ is the plasma pressure gradient length in the near SOL, are shown in the proximity of the separatrix for PT and NT plasmas. The shear rate decreases as the plasma collisionality increases since the high collisionality enhances the plasma transport, therefore flattening the pressure profile near the separatrix. While the normalized shear rate for NT plasma is found to be above one in all cases, it is not sufficiently strong to create a transport barrier or destabilize a Kelvin-Helmholtz (KH) instability \cite{Rogers2005, Myra2016}.

\begin{figure}[H]
\begin{center}
    {\includegraphics[width=0.8\textwidth]{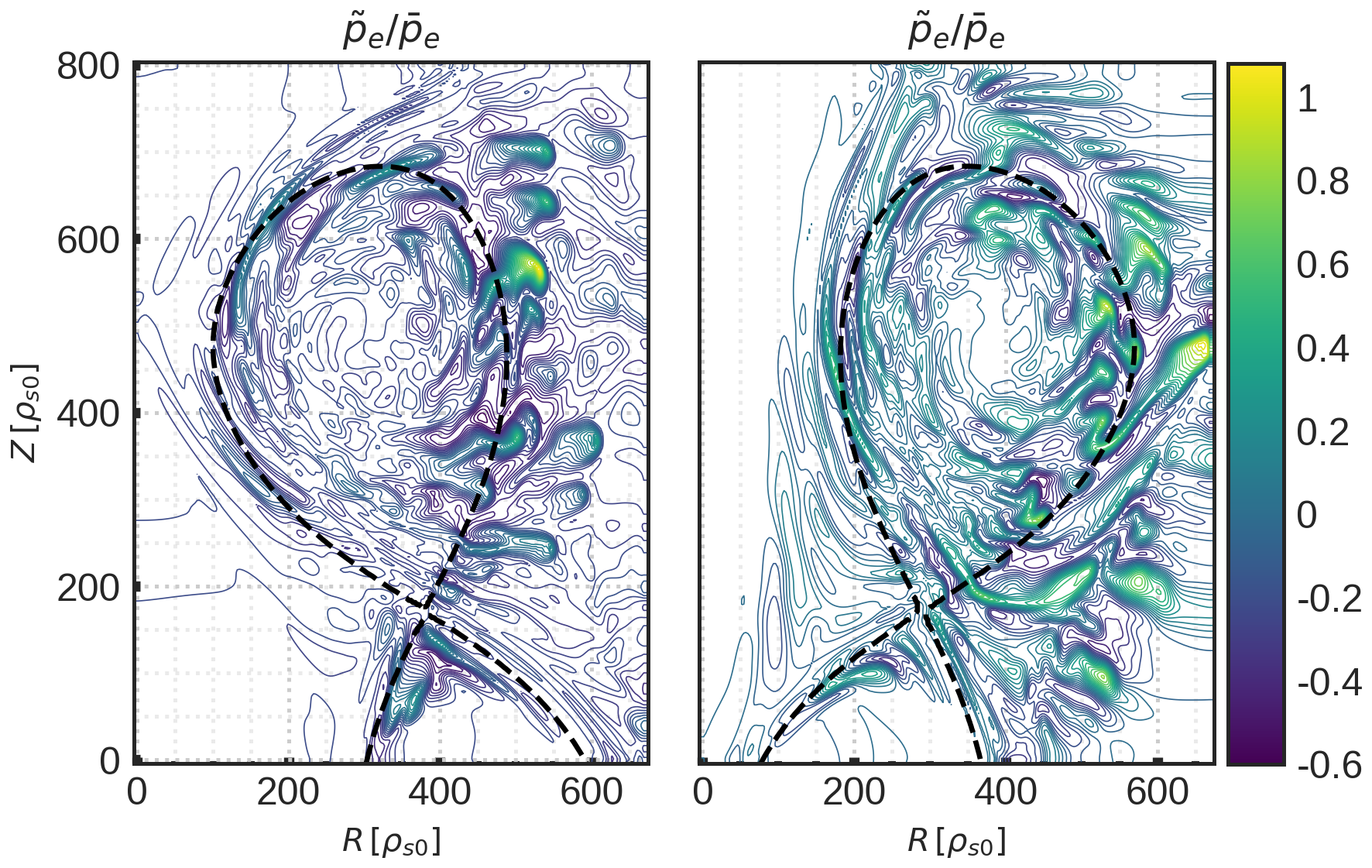}}
\caption{Snapshot of the fluctuating electron pressure normalized to the equilibrium electron pressure for a NT and a PT plasma with $s_{T0}=0.025$ and $\nu_0=1.0$.}
\label{2D_fluct_pe}
\end{center}
\end{figure}

Figure \ref{2D_fluct_pe} shows the contour of normalized electron pressure fluctuations. While small amplitude fluctuations are observed in the core, both NT and PT plasma display larger fluctuation amplitude in the edge region, as well as the presence of intermittent coherent structures, known as blobs \cite{Ippolito2011}, in the far SOL region. Blobs have larger size in the PT plasma. In agreement with previous results reported from simulations of limited configurations \cite{Riva2017}, NT plasmas are characterized by lower fluctuation levels and smaller eddy size compared to the PT counterpart. These results are in line with recent TCV experiments that point out a substantial reduction of density and temperature fluctuation amplitude and turbulence correlation length near the edge region in NT L-mode plasmas with respect to PT discharges \cite{Huang2019, Merlo2021}. 

\begin{figure}[H]
\begin{center}
    {\includegraphics[width=0.4\textwidth]{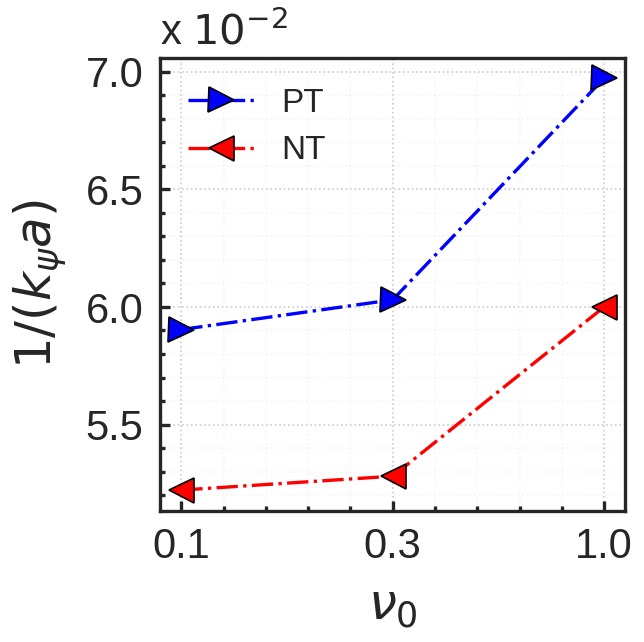}}
\caption{Normalized radial extension of the largest turbulent eddies at the LFS, $1/(k_\psi a)$ with $k_\psi$ being the radial wave number, averaged in time and along the toroidal direction.}
\label{kpsi}
\end{center}
\end{figure}

For a qualitative analysis of the turbulence eddy size, we measure the radial extension of the turbulent structure at the low-field side (LFS), $1/k_\psi$. This is defined as the distance where the cross-correlation drops to 0.5. The radial length is then averaged in time and along the toroidal direction. The results, normalized to the tokamak minor radius, $1/(k_\psi a)$, are shown in Figure \ref{kpsi} for PT and NT plasmas as a function of different values of collisionality. Confirming our qualitative analysis, the size of turbulent eddies in NT plasmas is smaller than PT plasmas. Furthermore, the radial size of the turbulent structure increases with plasma collisionality, indicating the presence of large scale turbulence at high value of $\nu_0$ \cite{Ippolito2011}. The typical values of $1/(k_\psi a)$ for different turbulent regimes are identified in Ref. \cite{Giacomin2020}, i.e. $1/(k_\psi a) \ll 1$ for RDWs and $1/(k_\psi a) \sim 0.1$ for RBMs, suggesting that the considered simulations are mainly governed by RBMs. This is confirmed by tests where we zero out the interchange drive, i.e. the curvature term in Eq. (\ref{GBS_vorticity}), and we observe a significant steepening of the pressure profile.

%% file: 4_Lp_estimate.tex
\section{Estimate of the pressure gradient length}\label{Sec4}
In this section, an analytical estimate of the plasma pressure gradient length in the near SOL, $L_p = -p/\nabla p $, is derived. This is correlated to the power fall-off length $\lambda_q$ that regulates divertor heat load on the outer target \cite{Loarte2007, Silvagni2020}. With the aim of predicting the SOL width as a function of the operational parameters, a theoretical scaling law based on first principle approach is derived for diverted configurations in Ref. \cite{Giacomin2021_2} leveraging previous work in limited configurations \cite{Halpern2013}. The scaling is validated against an experimental dataset for different tokamaks and nonlinear GBS simulations. We extend here the scaling obtained in Ref. \cite{Giacomin2021_2} to include the effect of triangularity and elongation. The derivation is based on a quasi-linear analysis where the gradient removal mechanism \cite{Ricci2013}, i.e. the local flattening of the plasma pressure profile, provides the main mechanism for the saturation of the growth of the linear instabilities driving turbulence. The value of $L_p$ is then obtained by a balance between perpendicular turbulent transport and parallel losses at the end of the magnetic field lines.

For the derivation, the flux coordinates $(x,y,z)$ introduced in Section \ref{Sec2} are used. As the radial flux in the edge plasma is mainly driven by turbulence, it can be estimated as $\Gamma_x \sim \overline{\tilde{p}_e \partial_y \tilde{\phi}}$. The relation between $\tilde{p}_e$ and $\partial_y \tilde{\phi}$ can be obtained from the linearized electron pressure equation by combining Eqs. (\ref{GBS_density}) and (\ref{GBS_electron_temperature}):
\begin{eqnarray}
    \gamma \tilde{p}_e &\sim -\frac{\rho_*^{-1}}{B}[\tilde{\phi}, \bar{p}_{e}]  \\
     &\sim \rho_*^{-1}\partial_y \tilde{\phi} \partial_x \bar{p}_{e},
\end{eqnarray}
where the curvature and parallel gradient terms are neglected by retaining only leading order contributions. The saturation of the growth of the instabilities occurs when their amplitude is sufficient to remove their driving gradient. In this condition, the radial gradient associated with the pressure fluctuations is comparable to the radial gradient of the background pressure, i.e. $\tilde{p}/\bar{p} \sim 1/(k_x L_p)$ where $k_x \sim \sqrt{k_y/L_p}$ provides an estimate of the radial eddy extension, according to a non-local linear theory \cite{Ricci2008}. This allows us to express the radial flux as
\begin{eqnarray}
    \Gamma_x \sim \rho_* \gamma \frac{\tilde{p_e}^2}{\bar{p}_e}L_p \sim \rho_* \frac{\gamma}{k_{x}^2} \frac{\bar{p}_e}{L_p} \sim \rho_* \frac{\gamma}{k_{y}} \bar{p}_e.
    \label{Heat_flux}
\end{eqnarray}

The balance between the perpendicular turbulent transport, $\partial_x \Gamma_x \sim \Gamma_x/L_p \sim \rho_*\bar{p}_e \gamma / (k_y L_p)$, and the parallel losses at the sheath, $\nabla_\parallel (p v_{\parallel e})\sim \rho_*\bar{p}_e c_s /q$, then leads to an estimate of the pressure scale length
\begin{equation}
    L_p \sim \frac{q}{c_s}\Bigg(\frac{\gamma}{k_y}\Bigg)_{\rm{max}}.
\label{Lp_lin2}
\end{equation}   

In order to evaluate $(\gamma/k_y)_{\rm{max}}$ in Eq. (\ref{Lp_lin2}), similarly to previous work \cite{Riva2017}, we linearize Eqs. (\ref{GBS_density}-\ref{GBS_electron_temperature}). For this purpose, all physical quantities are expressed as a sum of an equilibrium and a perturbation component, i.e. $n(x,y,z,t)=n_0(x)+\delta n(y,z,t)$, with $\delta n(y,z,t)=\delta n(z)\exp{(ik_y y + \gamma t)}$. By assuming a density equilibrium gradient, $\partial_x n = -n_0/L_n$, where $L_n=-n/\nabla n$ is the characteristic scale length, and making a similar assumption for $T_e$, while other equilibrium quantities are assumed to vanish ($\phi_0=v_{\parallel i,0}=v_{\parallel e,0}=0$) the linearized GBS system normalized to the separatrix value can be recast as:
\begin{eqnarray}
    \fl \gamma \delta n &= \frac{R_0}{L_n} \frac{1}{B}\mathcal{P}^L (\delta \phi) + \frac{2}{B}\mathcal{C}^L(\delta p_e - \delta \phi) + (\nabla_{\parallel} + \nabla \cdot \bm{b})(\delta j_\parallel - \delta v_{\parallel i}),\label{GBS_lin1} \\
    \fl\frac{1}{B^2} \gamma \delta{\omega} &=\frac{2}{B}\mathcal{C}^L(\delta p_e) + (\nabla_\parallel + \nabla \cdot \bm{b})\delta j_\parallel, \\
    \fl\frac{m_e}{m_i}\gamma \delta v_{\parallel e} &= \nabla_\parallel (\delta \phi - \delta p_e -0.71\delta T_e) + \nu \delta j_\parallel, \\
    \fl \gamma \delta v_{\parallel i} &= - \nabla_\parallel \delta p_e, \\
    \fl \gamma \delta T_e &= \frac{R_0}{L_n} \frac{\eta}{B}\mathcal{P}^L(\delta \phi) + \frac{4}{3B}\mathcal{C}^L \bigg( \delta p_e + \frac{5}{2}\delta T_e - \delta \phi \bigg) \nonumber \\
    &     + \frac{2}{3}(\nabla_\parallel + \nabla \cdot \bm{b})(1.71\delta j_{\parallel} - \delta \nu_{\parallel i}),\label{GBS_lin5}
\end{eqnarray}
where we define $\delta p_e = \delta n + \delta T_e$, $\delta j_\parallel = \delta v_{\parallel i} - \delta v_{\parallel e}$, $\delta \omega = (\nabla^2_\perp)^L \delta \phi$ and $\eta=L_n/L_{T_e}$. In addition, the linearized expressions of the geometrical operators in Eqs. (\ref{Operator_Poisson}-\ref{Operator_Laplacian}) can be simplified as
\begin{eqnarray}
    \mathcal{P}^L(f) &= \mathcal{P}_{xy}\frac{\partial f}{\partial y} + \mathcal{P}_{yz}\frac{\partial f}{\partial z}\simeq i\mathcal{P}^{xy}k_yf, \\ \mathcal{C}^L(f) &= \mathcal{C}^y \frac{\partial f}{\partial y} + \mathcal{C}^z \frac{\partial f}{\partial z} \simeq i\mathcal{C}^y k_y f, \\
    (\nabla_\perp^2)^Lf &= \mathcal{N}^{yy}\frac{\partial^2 f}{\partial y^2}+\mathcal{N}^y \frac{\partial f}{\partial y} \simeq -\mathcal{N}^{yy}k_y^2 f,
\end{eqnarray}
where we neglect the relatively small $\mathcal{P}^{yz}, \mathcal{C}^z$ and $\mathcal{N}^y$ terms (see \ref{Appendix:coef}).

Provided that turbulence in our nonlinear simulations is mainly driven by RBMs in the bad curvature region, we assume a strongly localized mode at $\theta=0$, $k_z \sim 1/q$ and $\epsilon=0$ to simplify Eqs. (\ref{GBS_lin1}-\ref{GBS_lin5}). The analytical expressions of $\gamma$ and $k$ is then obtained by evaluating $\partial_{k_y}(\gamma/k_y)_{\rm{max}}=0$ as a function of the shaping parameters \cite{Riva2017}. As a result, the linear growth rate $\gamma$ and the poloidal wavenumber $k_y$ are given by 
\begin{eqnarray}
    \gamma^2 &= \gamma_{\rm{RBM}}^2 \frac{\mathcal{C}(\kappa, \delta, q)}{3},\label{gamma} \\
    k_y^2 &= \frac{\sqrt{3}}{2} k_{\rm{RBM}}^2 \mathcal{C}(\kappa, \delta, q)^{-1/2},\label{kchi}
\end{eqnarray}
with $\gamma_{\rm{RBM}} = \sqrt{(2
\bar{T}_e)/(\rho_* L_p)}$ and $k_{\rm{RBM}} = 1/\sqrt{\bar{n}\nu q^2 \gamma_{\rm{RBM}}}$, while the effect of $\kappa$ and $\delta$ is contained in the curvature coefficient $\mathcal{C}$. This is evaluated by using the fact that RBMs are mostly destabilized at the LFS. The analytical expressions of $\mathcal{C}$ at the outer midplane ($\theta=0$) can then be approximated in the large aspect ratio limit (see \ref{appendix:curvature})

\begin{equation}
    \fl\mathcal{C}(\kappa, \delta, q) = \frac{\partial R_c(r,\theta)}{\partial r}\Biggr|_{{\theta=0}} = 1-\frac{\kappa-1}{\kappa+1}\frac{3q}{q+2} + \frac{\delta q}{1+q} + \frac{(\kappa-1)^2(5q-2)}{2(\kappa+1)^2(q+2)} + \frac{\delta^2}{16}\frac{7q-1}{1+q}.
    \label{Curvature_operator}
\end{equation}

\begin{figure}[H]
\begin{center}
\includegraphics[width=0.48\textwidth]{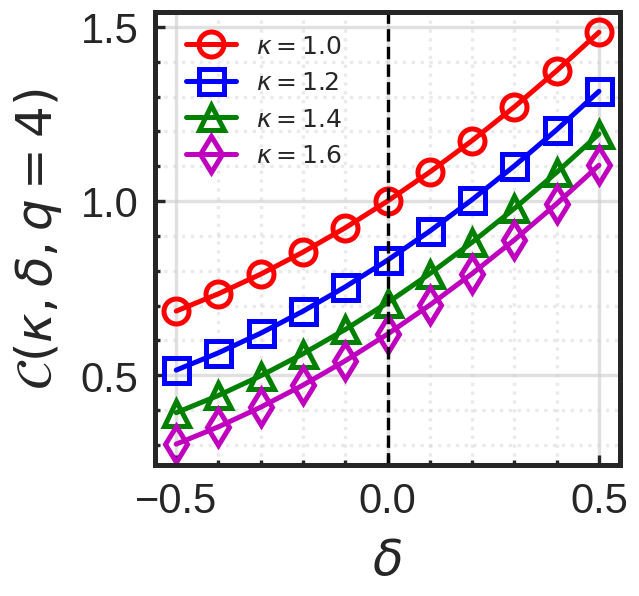}
\caption{Curvature coefficient in Eq. (\ref{Curvature_operator}), $\mathcal{C}(\kappa, \delta, q)$, at the outer midplane as a function of $\delta$ for different values of $\kappa$ with $q=4$. The case of $\kappa=1.0$ and $\delta=0$ corresponds to the circular plasma, yielding $\mathcal{C}(\kappa, \delta, q)=1$.}
\label{Ch4:Curvature}
\end{center}
\end{figure}
In Figure \ref{Ch4:Curvature}, the value of the curvature coefficient $\mathcal{C}(\kappa, \delta, q)$ in Eq. (\ref{Curvature_operator}) is shown as a function of $\kappa$ and $\delta$, while considering a safety factor constant  $(q=4)$. Similarly to previous results \cite{Riva2017}, $\delta$ is found to play a more important role than $\kappa$ in determining the value of $\mathcal{C}(\kappa, \delta, q)$. A similar reduction of the curvature coefficient with $\delta$ is visible also in the GBS curvature operator in \ref{Appendix:coef}.

We can now predict the value of the pressure gradient length $L_p$ as a function of the tokamak operational parameters. We consider the balance between the heat fluxes crossing the LCFS and the volume of integrated heat source,
\begin{equation}
    S_p(R,Z) \simeq \oint_{LCFS} q_{x}(R,Z)dl \sim L_\chi q_{x,i},
    \label{Volume_integration_heat}
\end{equation}
where $S_p(R,Z)=\int s_p (R,Z)dRdZ$ with $s_p=ns_{T_e}+T_e s_n$, and $L_\chi=\oint_{LCFS}dl$ the poloidal length of the LCFS. A common approach to obtain $L_\chi$ is to simply approximate it to the circumference of ellipse \cite{Sauter2016}. Considering that transport in the RBM regime mostly occurs at the LFS, this gives, i.e. $L_\chi \simeq \pi a \sqrt{(1+\kappa^2)/2}$. In the case of a triangular plasma, the assumption of approximating $L_\chi$ to the circumference of an ellipse leads to an error of over 10\% for $\bigr|\delta\bigr|>0.5$. To address this issue, we modify the expression for $L_\chi$ by numerically computing the poloidal length and apply a Taylor expansion around $\kappa=1$ and $\delta=0$. The resulting expression accounts for the effect of triangularity and can be expressed as
\begin{equation}
    L_\chi \simeq \pi a (0.45 + 0.55\kappa) + 1.33 a \delta,
\label{Lchi_analytic}
\end{equation}
where, the error with respect to the numerical values is found to be less than $3\%$ when $\kappa = 1$ and $\bigr|\delta\bigr|<0.5$.

Finally, by equating Eqs. (\ref{Heat_flux}) and (\ref{Volume_integration_heat}), the analytical estimate of $L_p$, including effects of plasma shaping, can be recast as: 
\begin{eqnarray}
    L_{p} \sim \mathcal{C}(\kappa, \delta, q) \bigg[ \rho_* (\nu_0 \bar{n} q^2)^2 \bigg(\frac{L_\chi \bar{p}_e}{S_p}\bigg)^{4} \bigg]^{1/3},
    \label{Lp_analytical}
\end{eqnarray}
where $\mathcal{C}$ is the poloidal curvature coefficient defined in Eq. (\ref{Curvature_operator}). Note that the above equation is equivalent to the scaling derived in Ref. \cite{Giacomin2020} when the shaping term $\mathcal{C}(\kappa, \delta, q)$ is neglected.

\begin{figure}[H]
\begin{center}
    \includegraphics[width=\textwidth]{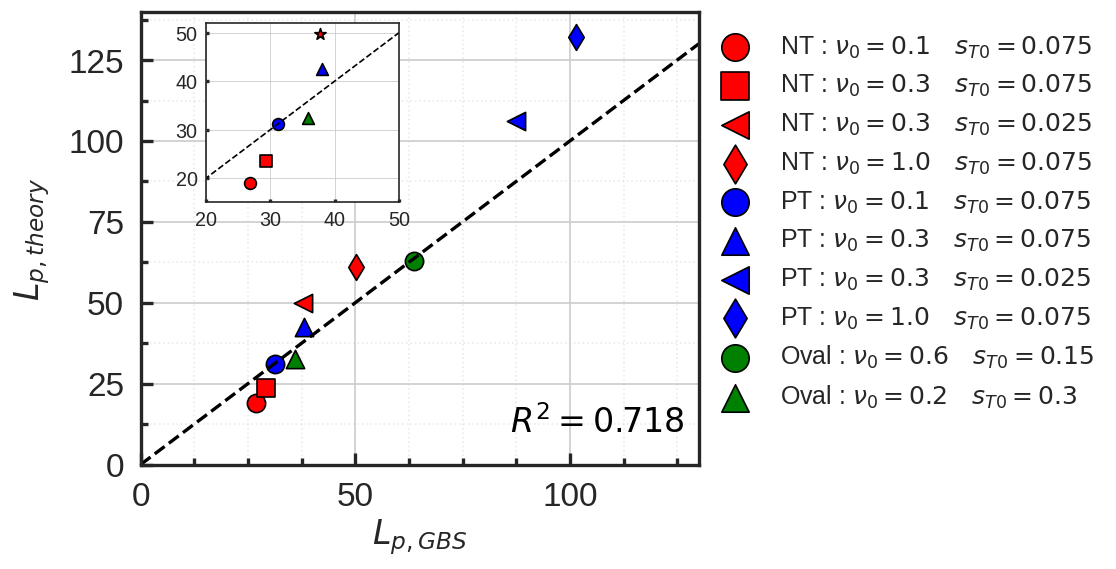}
\caption{Comparison of the pressure gradient length $L_p$ between the analytical scaling law in Eq. (\ref{Lp_analytical}) and the value of $L_p$ obtained from nonlinear GBS simulation. A parametric scan for collisionality $\nu_0$ and heating power $s_{T0}$ is carried out for different plasma shapes. Here, 'Oval' denotes elongated but non-triangular $(\delta=0)$ plasma.}
\label{Ch4:Lp_NoBou}
\end{center}
\end{figure}

In Figure \ref{Ch4:Lp_NoBou}, the $L_p$ estimates provided by the analytical scaling law in Eq. (\ref{Lp_analytical}) is compared with the results of the nonlinear GBS simulations presented in Section \ref{Sec3}, for different values of $\nu_0$ and $s_{T0}$. Three different values of triangularity are considered, $\delta = -0.3, 0, 0.3$. Two remarks can be made from the observation of Figure \ref{Ch4:Lp_NoBou} where good agreement is observed between the simulations and analytical results, as shown by the high R-square factor ($R^2 = 0.718$). First, when different values of $\delta$ with fixed $\nu_0$ and $s_{T0}$ are compared, NT plasmas tend to yield smaller values of $L_p$, mainly because of the reduction of the curvature drive. Second, when different values of $\nu_0$ and $s_{T0}$ with fixed $\delta$ are compared, we observe that the size of $L_p$ increases with $\nu_0$ and decreases with $s_{T0}$, being $L_p$ related to the size of the turbulent eddies, which increases with the plasma resistivity \cite{Giacomin2020}.

%% file: 5_Lp_comparison.tex
\section{Comparison of \texorpdfstring{$L_p$}{Lg} with experimental data}\label{Sec5}
To validate the reliability of the pressure gradient scaling law derived in Section \ref{Sec4}, a comparison against an experimental database is performed. This leverages the work in Ref. \cite{Giacomin2021_2}, where a comparison between a scaling of the pressure decay length and the experimentally measured power fall-off length $\lambda_q$ is described, showing good agreement. The database considers discharges from MAST, TCV, JET, COMPASS and Alcator C-Mod tokamaks with triangularity $0.1 < \delta < 0.5$. The scaling law considered in Ref. \cite{Giacomin2021_2} does not take into account the effect of triangularity, and can be obtained from Eq. (\ref{Lp_analytical}) by neglecting the shaping term. However, recent works with TCV and ASDEX Upgrade have experimentally shown that the power fall-off length at the outer target in L-mode plasmas varies by a factor of two in NT plasmas compared to PT discharges \cite{Faitsch2018}.

\begin{figure}[H]
\begin{center}
\includegraphics[width=\textwidth]{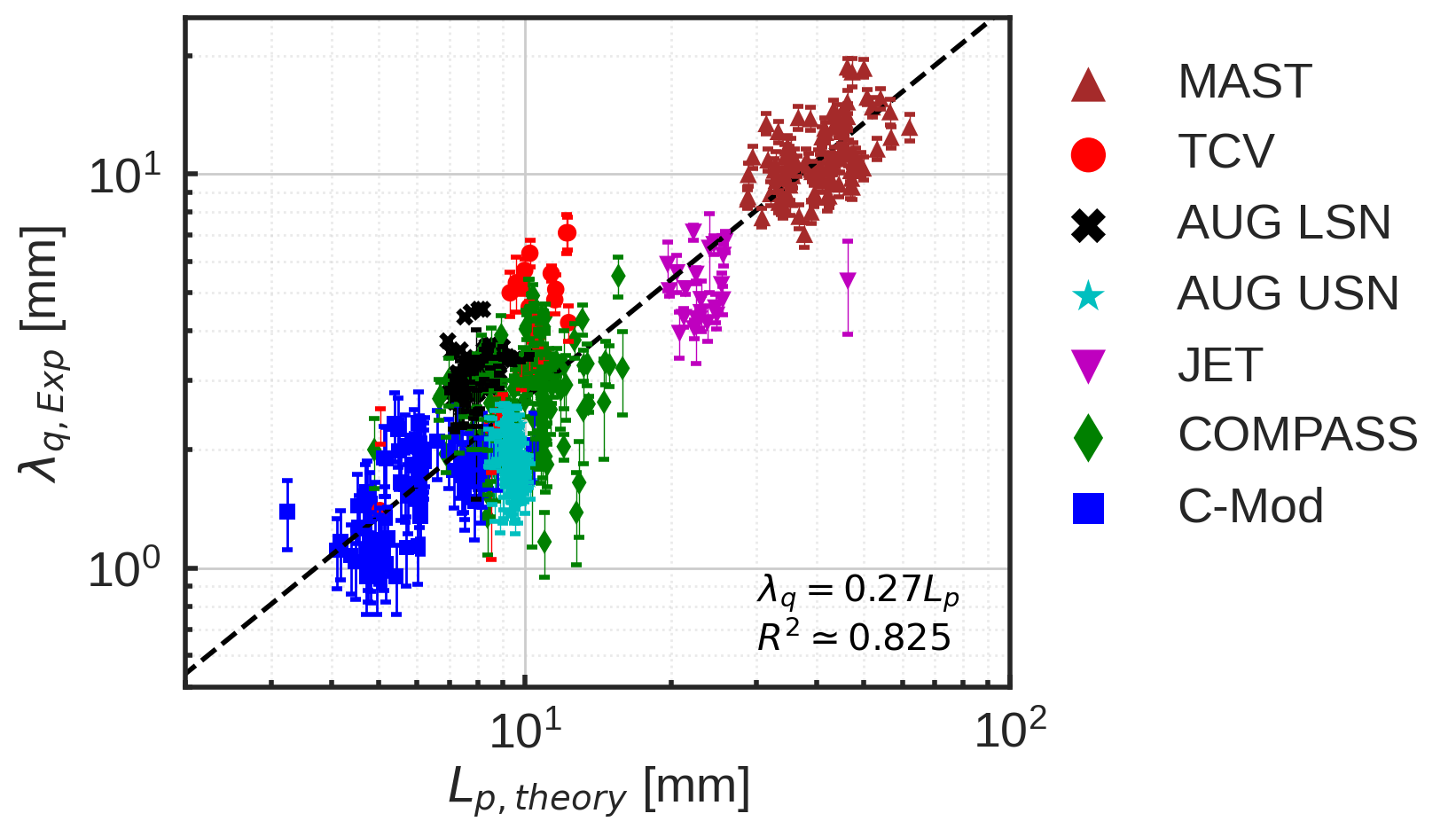}
\caption{Comparison between the theoretical pressure gradient scaling law for  $L_p$ by Eq. (\ref{Lp_engineering}), and experimental power fall-off length $\lambda_q$ from multi-machine database. The fitting coefficient ($\alpha=0.27$) is obtained using a least-squares method. We obtain a fitting quality $R^2 \simeq 0.825$.}
\label{Ch4:Lp_exp_comparison}
\end{center}
\end{figure}

For carrying out a comparison of the scaling in Eq. (\ref{Ch4:Lp_exp_comparison}) with experimental results, we make use of the experimental dataset already considered for the validation in Ref. \cite{Giacomin2021_2}. This dataset is described in Ref. \cite{Horacek2020}, and includes a set of power fall-off decay lengths measured from both Langmuir probes or infrared cameras in the MAST, JET, COMPASS and Alcator C-Mod tokamaks. The discharges considered in the database are single-null L-mode plasmas in attached conditions where the pressure gradient between upstream and target is negligible. This database is expanded here by including the ASDEX Upgrade data described in Ref. \cite{Silvagni2020}. The ASDEX Upgrade discharges include lower single-null (LSN) configurations, favourable for H-mode access, and upper single null (USN) unfavorable configurations, performed in L-mode plasmas, with triangularity $0.1 < \delta < 0.3$. In addition, we include recent TCV discharges dedicated to the study of triangularity effects on L-mode plasmas, extending our range of triangularity to $-0.3 < \delta < 0.5$.

For a direct comparison with experimental data, the scaling law in Eq. (\ref{Lp_analytical}) is rewritten in terms of engineering parameters. By applying the same procedure as in Ref. \cite{Giacomin2021_2}, we express $S_p$ as $P_{\rm{SOL}}/(2\pi R_0)$ and $\nu_0$ using Eq. (\ref{Collisionality}). Then Eq. (\ref{Lp_analytical}) results into the following expression
\begin{eqnarray}
    \fl L_{p} \simeq 1.95 \, \mathcal{C}(\kappa, \delta, q)^{9/17} A^{1/17} q^{12/17} R_0^{7/17}P_{\rm{SOL}}^{-4/17}n_e^{10/17}B_T^{-12/17}L_\chi^{12/17},
\label{Lp_engineering}
\end{eqnarray}
where we use physical units, i.e. $L_p$[mm], $P_{\rm{SOL}}$[MW], $R_0$[m], $a$[m], $n_e$[$10^{19}\rm{m}^{-3}$] and $B_T$[T]. Note that effects of plasma shaping are included in both $\mathcal{C}(\kappa, \delta, q)$ and $L_\chi$ terms (see Eqs. (\ref{Curvature_operator}) and (\ref{Lchi_analytic})).

In Figure \ref{Ch4:Lp_exp_comparison}, we show the result of the comparison of the scaling law in Eq. (\ref{Lp_engineering}) with the experimentally measured $\lambda_q$ at the outer target for the entire multi-machine database. Overall, very good agreement is observed with very high quality of fitting, $R^2 \simeq 0.825$ with the root-mean-square (RMS) error $\pm 1.4$mm, where the proportionality constant $\alpha=0.27$ is used.
The newly derived scaling law for $L_p$ is found to produce a slightly better fitting result compared to the previous scaling law derived in Ref. \cite{Giacomin2021_2}, which did not include the dependence on triangularity with a value of $R^2 \simeq 0.807$. In particular, the scaling law we propose provides better estimates for the MAST discharges \cite{Sykes2001}, which is characterized by strongly shaped plasmas, approximately with $\kappa \sim 2$ and $\delta \sim 0.5$, showing larger values of $L_p$ compared to other tokamaks. 

The experimental data used for the scaling law in Figure \ref{Ch4:Lp_exp_comparison} are found to be either in the sheath-limited regime or in the weak conduction-limited regime. This is verified by evaluating the SOL collisionality, $\nu^*=10^{-16}n_u L /T_u^2$, for all discharges. The results reveal that the majority of data have $3< \nu^*<10$, except for a few discharges from COMPASS, TCV and ASDEX Upgrade in the USN configuration discharges, characterized by $10<\nu^*<22$. Note that the ASDEX Upgrade data for LSN (favorable) and USN (unfavorable) configurations show the effect of the toroidal magnetic field direction \cite{Faitsch2015} with a larger SOL collisionality for USN ($10<\nu^*<13$) than the LSN discharges ($\nu^*<10$), yielding a slight deviation from pure sheath-limited regime for USN discharges. This effect can play a role in the observed difference between LSN and USN discharges in Figure \ref{Ch4:Lp_exp_comparison}, and will be a subject for future work.

\begin{figure}[H]
\begin{center}
	\subfloat[NT $(\delta=-0.3)$]{\includegraphics[width=0.25\textwidth]{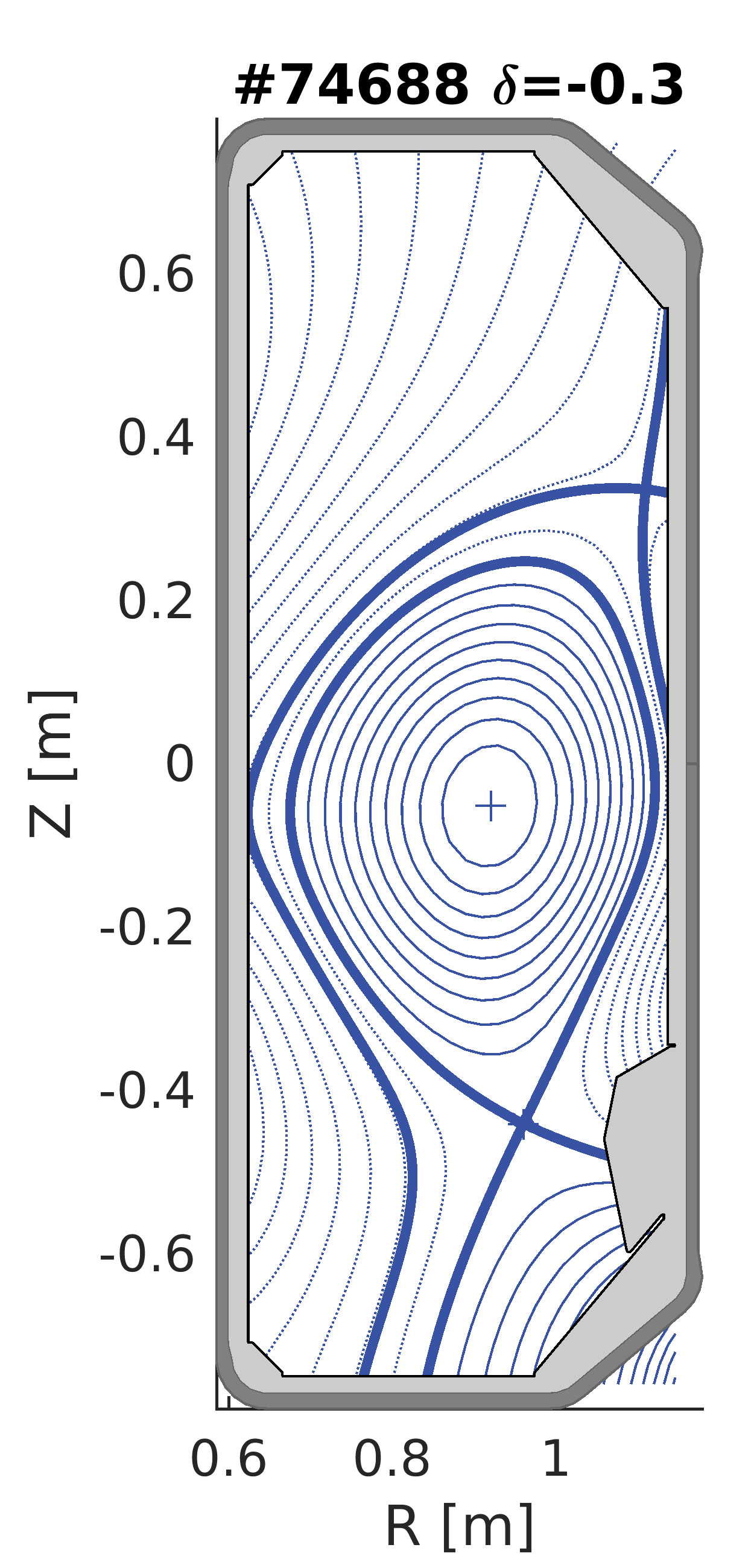}}\label{TCV_NT}
	\subfloat[PT $(\delta=+0.15)$]{\includegraphics[width=0.25\textwidth]{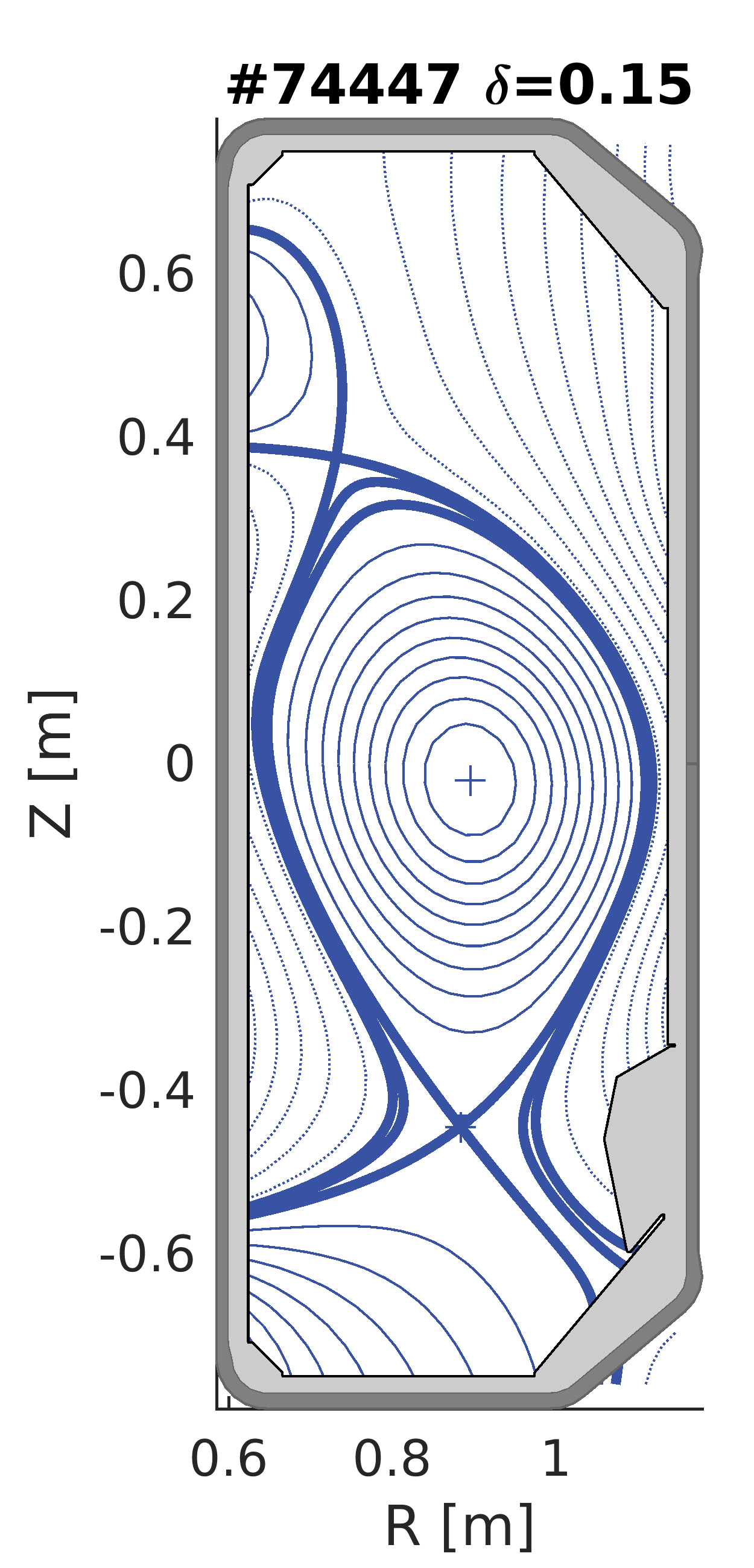}}\label{TCV_PT}
\caption{Poloidal cross section of TCV magnetic equilibria for the NT $(\#74688)$ and PT $(\#74447)$ discharges.}
\label{Ch4:TCV_equilibria}
\end{center}
\end{figure}

As a further validation of our scaling law, we consider two L-mode TCV discharges with NT and PT configurations. While keeping other TCV parameters approximately constant ($B_T = 1.43T$, favorable ion-$\nabla B$ drift direction, the plasma current $I_p = 220\textrm{kA}$, line-averaged density $\langle n_e \rangle = 2.2-2.6 \times 10^{19}m^{-3}$ and $\kappa=1.5$), the value of $\delta$ is varied to investigate the effect of triangularity on the outer divertor target heat flux. In particular, we consider two discharges, which features NT $(\delta=-0.3)$ and PT $(\delta=0.15)$ magnetic equilibria, whose flux surfaces are depicted respectively in Figure \ref{Ch4:TCV_equilibria}. The line-averaged plasma density and current measurements indicate that the discharges under consideration are operating within the attached regime.

\begin{figure}[H]
\begin{center}
	\includegraphics[width=0.8\textwidth]{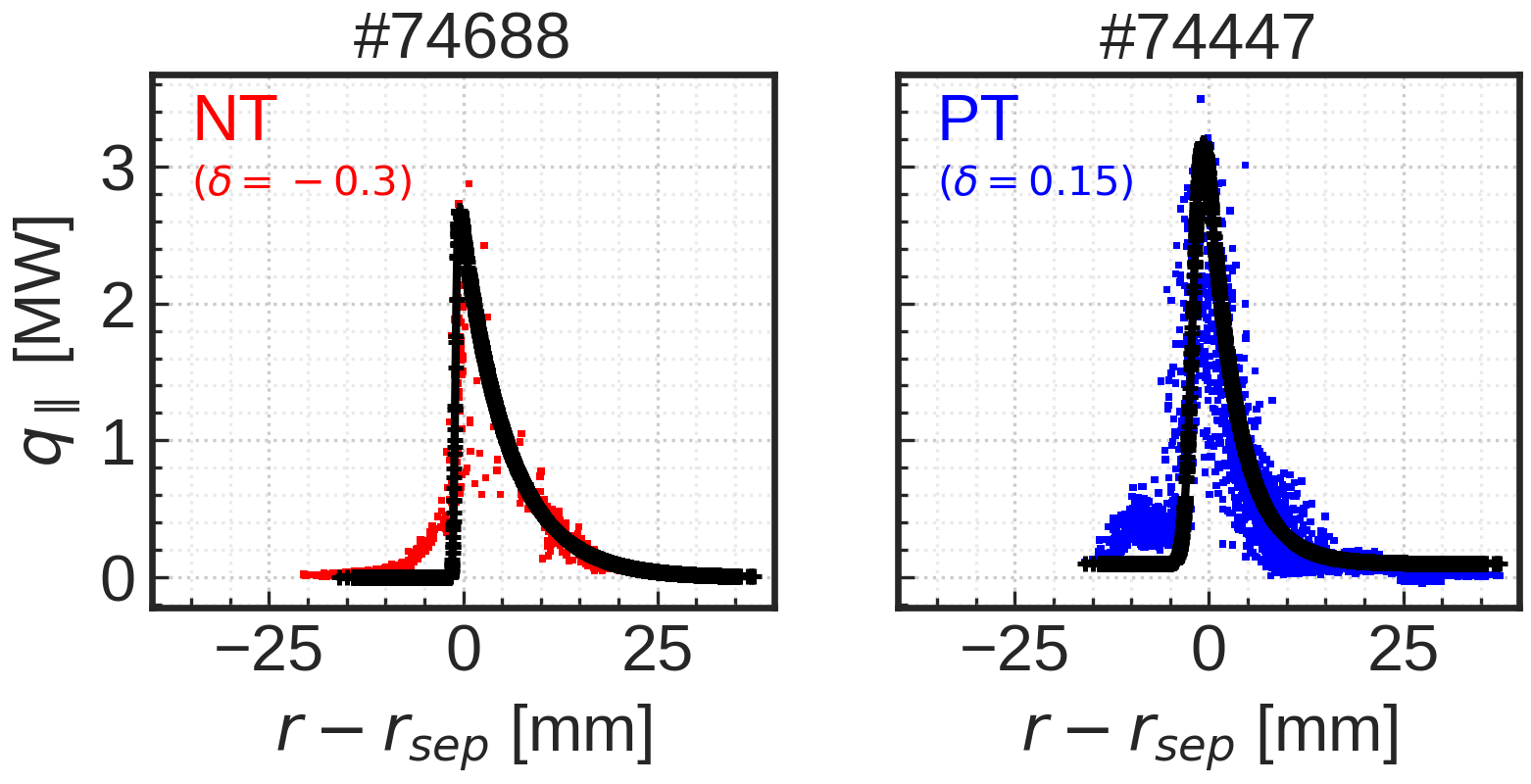}
\caption{Parallel heat flux of the outer target remapped to the outer midplane. By using the Eich-fit \cite{Eich2013} (black dashed line), the size of $\lambda_q$ is found to be $\sim 3.6\rm{mm}$ for NT and $\sim 4.7\rm{mm}$ for PT plasma.}
\label{Ch4:TCV_qpar}
\end{center}
\end{figure}

In Figure \ref{Ch4:TCV_qpar}, the TCV outer target heat flux, measured from Langmuir probes \cite{Fevrier2018, Oliveira2019, Gorno2023}, remapped to the outer midplane is shown for both discharges using the Eich-fit (black solid line) \cite{Eich2013} to evaluate $\lambda_q$. The peak parallel heat-flux is reduced up to $\sim 30\%$ in the case of the NT plasma. The experimental values of $\lambda_q$ for NT and PT plasmas measured at the outer target result in $\lambda_{q,\rm{NT}} \simeq 3.6 \pm 0.55 \rm{mm}$ and $\lambda_{q,\rm{PT}} \simeq 4.7 \pm 0.93 \rm{mm}$ respectively. The $L_p$ scaling law in Eq. (\ref{Lp_analytical}) predicts $\lambda_{q,NT}\simeq 1.4 \pm 0.24\rm{mm}$ and $\lambda_{q,PT} \simeq 2.3 \pm 0.35\rm{mm}$, using the proportionality constant $\alpha=0.27$ found in the analysis of the multi-machine database. The theoretical scaling law reproduces the increase of $\lambda_q$ proportional to $\delta$, with an error comparable to the RMS error found in the analysis of the multi-machine database (we note that, in the analysis multi-machine in Figure \ref{Ch4:Lp_exp_comparison}, larger experimental $L_p$ than the theoretical scaling are observed for the TCV tokamak). The theoretical prediction we provide is accompanied by an error bar that reflects the uncertainty in the power entering the SOL, specifically the radiative power. Furthermore, the difference between the scaling law and the experimental measurement can be attributed to the use of the global value of $\delta$, instead of distinguishing between $\delta_{\rm{upper}}$ and $\delta_{\rm{lower}}$ \cite{Faitsch2018}.

%% file: 6_Conclusion.tex
\section{Conclusions}\label{Sec6}
In the present paper, the effects of triangularity on boundary plasma turbulence are investigated using global, flux-driven, two-fluid GBS simulations. A first-principles theoretical scaling law of the SOL width including triangularity is derived based on considerations of the linear growth rate and nonlinear saturation mechanisms of the driving instabilities. Overall, plasma shaping parameters are found to be important elements to determine the properties of boundary plasma turbulence, in particular the power fall-off length $\lambda_q$. 

A series of nonlinear GBS simulations is carried out for NT and PT L-mode diverted plasmas. NT plasmas show a stabilizing effects on edge plasma turbulence yielding (i) higher electron equilibrium pressure, (ii) reduced eddy size, (iii) steeper plasma gradient at the separatrix and (iv) improved energy confinement, with respect to PT simulations. Turbulence stabilization is due to the reduction of the magnetic curvature drive. Indeed, the curvature operator is found to decrease at the LFS for an elongated NT plasma resulting into stabilized SOL plasma turbulence, especially when RBM is the driving instability.

Leveraging the analysis of the simulation results and as an extension of the previous work in Ref. \cite{Giacomin2021_2}, a theoretical scaling law for $L_p$ is derived to include effect of triangularity in the $L_p$ estimate. The scaling law is then compared to the results of GBS simulations. The linear analysis shows weak growth rate and higher value of poloidal wavenumber $k_y$ (poloidally decorrelated turbulence structure) in an elongated NT plasma. The comparison of the theoretical scaling with results of a set of GBS simulations at different collisionality and input power shows an overall good agreement, with the analytical scaling, correctly capturing the $L_p$ reduction observed in NT GBS simulations. The above results are consistent with the experimental work reported in Ref. \cite{Faitsch2018}.

Finally, a comparison between the theory-based $L_p$ scaling law and experimental $\lambda_q$ dataset from different tokamaks is successfully performed. The derived scaling law can be used to predict the SOL width in future devices. In Table \ref{Extrapolation}, the SOL power fall-off length $\lambda_q$ predictions provided by Eq. (\ref{Lp_engineering}) for future tokamaks, such as ITER \cite{Campbell2016}, DTT \cite{Contessa2019}, SPARC \cite{Fernandez2022} and JT-60SA \cite{Giruzzi2017}, are listed assuming operations in L-mode and a triangularity value based on their baseline scenario. These tokamaks consider baseline scenarios operation in a PT configuration. As a comparison, we also report the expected $\lambda_q$ value in NT, with opposite triangularity than the baseline scenario. In particular, we note that the predicted $\lambda_q$ in ITER for NT L-mode plasma yields $\lambda_{q} \simeq 2$mm, which is twice as large as the predicted $\lambda_{q} \sim 1$mm for the H-mode burning plasma scenario \cite{Eich2013}, highlighting the attractiveness of NT L-mode plasma in terms of handling the exhaust of the divertor targets, in addition to the advantage of operating ELM-free scenarios \cite{Marinoni2021}.      

\begin{table}[H]
\begin{center}
\begin{tabular}{c|c c c c} 
\hline\hline
Parameter  & ITER & DTT & SPARC & JT-60SA\\ \hline
$R_0$ [m]  & 6.2 & 2.1 & 1.85 & 2.96 \\
$a$ [m] & 2 & 0.6 & 0.57 & 1.18\\ 
$q_{95}$ & 3 & 3 & 3 & 3\\
$\kappa$ & 1.85 & 1.7 & 1.97 & 1.95\\
$\delta$ & 0.49 & 0.3 & 0.54 & 0.53\\
$\bar{n}_e$ $[\textrm{m}^{-3}]$ & $4 \times 10^{19}$ & $1.8 \times 10^{20}$ & $3.1 \times 10^{20}$ & $6.3 \times 10^{19}$\\
$B_T$ [T]& $5.3$ & $6$ & $12.2$ & 2.3\\
$P_{\rm{SOL}}$ [MW] & 18   & 15 & 29 & 10\\
\hline
$\lambda_{q,PT}$ [mm] & $\sim$4.7 & $\sim$2.6 & $\sim$2.1 & $\sim$ 6.8\\
$\lambda_{q,NT}$ [mm] & $\sim$2 & $\sim$1.5 & $\sim$0.8 & $\sim$ 2.6\\
\hline\hline
\end{tabular}
\end{center}
\caption{Power fall-off length extrapolation of future tokamaks for NT and PT L-mode plasmas. The values of $\lambda_{q,\rm{NT}}$ are computed using an opposite value of triangularity, $-\delta$, in the scaling law.}
\label{Extrapolation}
\end{table}

%% file: 7_Acknowledgements.tex
\section*{Acknowledgements}
The authors acknowledge T. Eich as well as the participants of the Festival de Th\'eorie 2022, Aix-en-Provence, for helpful discussions. 
This work has been carried out within the framework of the EUROfusion Consortium, partially funded by the European Union via the Euratom Research and Training Programme (Grant Agreement No 101052200 — EUROfusion). The Swiss contribution to this work has been funded by the Swiss State Secretariat for Education, Research and Innovation (SERI). Views and opinions expressed are however those of the author(s) only and do not necessarily reflect those of the European Union, the European Commission or SERI. Neither the European Union nor the European Commission nor SERI can be held responsible for them. The simulations presented herein were carried out in part on the CINECA Marconi supercomputer under the TSVV-1 and TSVV-2 project and in part at CSCS (Swiss National Supercomputing Center).

%% file: 8_Appendix_1.tex
\section{Geometrical coefficients and differential operators}\label{Appendix:coef}
We deduce the expressions of the geometrical coefficients presented in Eqs. (\ref{Operator_Poisson}-\ref{Operator_Laplacian}). Introducing the cylindrical coordinate system $\bm{R}=(R_c(r,\theta), \varphi_c, Z_c(r,\theta))$, the covariant metric tensor in the $(r,\theta, \varphi)$ coordinates are given by
\begin{eqnarray}
    g_{rr} &= \bigg( \frac{\partial R_c}{\partial r}\bigg)^2 + \bigg( \frac{\partial Z_c}{\partial r}\bigg)^2, \label{grr}\\
    g_{\theta r} &= \frac{\partial R_c}{\partial r}\frac{\partial R_c}{\partial \theta} + \frac{\partial Z_c}{\partial r}\frac{\partial Z_c}{\partial \theta}, \\
    g_{\theta \theta} &= \bigg( \frac{\partial R_c}{\partial \theta}\bigg)^2 + \bigg( \frac{\partial Z_c}{\partial \theta}\bigg)^2, \\
    g_{\varphi \varphi} &= R_c^2, \\
    g_{r\varphi} &= g_{\theta \varphi} = 0, \label{grphi}
\end{eqnarray}
where we use the definition of the metric coefficients 
\begin{equation}
    g_{ij}=\bm{e}_i \cdot \bm{e}_j = \frac{\partial \bm{R}}{\partial \xi_i} \cdot \frac{\partial \bm{R}}{\partial \xi_j},
\end{equation}
with $\bm{\xi}=(r, \theta, \varphi)=(\xi_1, \xi_2, \xi_3)$.

The contravariant metric tensors in $(r,\theta, \varphi)$ coordinates are obtained by inverting the covariant metric tensors. By using the relation given by
\begin{eqnarray}
g^{ij}=g^{mn}\frac{\partial u_i}{\partial u^m}\frac{\partial u^j}{\partial u^n},
\end{eqnarray}
the contravariant metric tensor in flux-tube coordinates $(r, \alpha, \theta_*)$, where $\alpha=\varphi-q(r)\theta_*$ is a field line label and $\theta_*$ is the straight-field-line angle defined in Eq. (\ref{field_line_angle}), can then be obtained. The resulting expressions can be recast as
\begin{eqnarray}
    g^{\theta_* \theta_*} &=\bigg( \frac{\partial \theta_*}{\partial \theta}\bigg)^2 g^{\theta \theta} +2\frac{\partial\theta_*}{\partial \theta}\frac{\partial \theta_*}{\partial \theta}g^{\theta r} + \bigg(\frac{\partial \theta_*}{\partial r}\bigg)^2 g^{rr}, \label{grr2} \\
    g^{\theta_* r} &= \frac{\partial \theta_*}{\partial r}g^{rr} + \frac{\partial \theta_*}{\partial \theta} g^{\theta r}, \\
    g^{r\theta_*} &=\frac{\partial \theta_*}{\partial r}g^{rr} + \frac{\partial \theta_*}{\partial \theta}g^{r\theta}, \\
    g^{\theta_* \alpha} &= -s(r) \theta_* \frac{q(r)}{r}g^{\theta_* r} - q(r) g^{\theta_* \theta_*}, \\
    g^{r \alpha} &= -s(r) \theta_* \frac{q(r)}{r}g^{rr}-q(r)g^{\theta_* r}, \\
    g^{\alpha \alpha} &= g^{\varphi \varphi} + q(r)^2 g^{\theta_*\theta_*} + 2\frac{q(r)^2 s(r) \theta_*}{r}g^{\theta_* r} + [s(r)\theta_*]^2 \frac{q(r)^2}{r^2} g^{rr},\label{grr3}
\end{eqnarray}
where $s(r)=(r/q)(dq/dr)$ is the magnetic shear.

By using the analytical expressions of the metric tensors defined in Eqs. (\ref{grr2}-\ref{grr3}), the geometrical coefficients presented in Eqs. (\ref{Operator_Poisson}-\ref{Operator_Laplacian}) can be analytically expressed in the re-scaled flux-tube coordinate system $x=r, y=(a/q)\alpha, z=qR_0\theta_*$ leading to
\begin{eqnarray}
    \fl\mathcal{P}_{xy}=-\frac{b_{\theta_*}a}{\mathcal{J}q}, \quad \mathcal{P}_{yz}=-\frac{ab_r}{\mathcal{J}}, \quad \mathcal{P}_{zx}=-\frac{qb_\alpha }{\mathcal{J}}, \label{coef_1} \\
    \fl\mathcal{D}^x=\mathcal{D}^y=0, \quad \mathcal{D}^z=qR_0b^{\theta_*},\\
    \fl\mathcal{C}^x=-\frac{R_0B}{2\mathcal{J}}\frac{\partial c_\alpha}{\partial \theta_*}, \quad \mathcal{C}^y=\frac{aR_0B}{2\mathcal{J}q}\bigg(\frac{\partial c_r}{\partial \theta_*}-\frac{\partial c_{\theta_*}}{\partial r}\bigg), \quad \mathcal{C}^z=\frac{qR_0B}{2\mathcal{J}}\frac{\partial c_\alpha}{\partial r}, \label{eq_curv}  \\
    \fl\mathcal{N}^{xx}=g^{rr}, \quad \mathcal{N}^{xy}=\frac{2g^{\alpha r}a}{q}, \quad \mathcal{N}^{yy}=\frac{a^2 g^{\alpha \alpha}}{q^2}, \\
    \fl\mathcal{N}^x=\nabla^2 r,\quad \mathcal{N}^y=\frac{a}{q}\nabla^2\alpha,\quad \mathcal{N}^z=qR_0\Big(\nabla^2 \theta_*-\frac{1}{\mathcal{J}}\frac{\partial}{\partial \theta_*}[\mathcal{J}(b^{\theta_*})^2]\Big), \\
    \fl\mathcal{N}^{xz}=2q g^{r\theta_*}, \quad \mathcal{N}^{yz}=2 a g^{\theta_* \alpha}, \quad \mathcal{N}^{zz}=q^2 [g^{\theta_* \theta_*}-(b^{\theta_*})^2],\label{coef_2}
\end{eqnarray}
where $c_i = b_i/B$.

Due to the fact that the scale length of the turbulence along the radial direction is larger than along the poloidal direction $(k_y \gg k_x)$ and that the parallel turbulence wavelengths is such that $(k_z \ll 1)$, the operators in Eqs. (\ref{Operator_Poisson}-\ref{Operator_Laplacian}) can be further simplified. For example, the curvature operator in flux-tube coordinates can be written as
\begin{eqnarray}
    \mathcal{C}(f) &= \mathcal{C}^x \frac{\partial f}{\partial x} + \mathcal{C}^y \frac{\partial f}{\partial y} + \mathcal{C}^z \frac{\partial f}{\partial z} \nonumber\\
    &\simeq \mathcal{C}^y \frac{\partial A}{\partial y}. \label{appendix:curvature2}
\end{eqnarray}

%% file: 8_Appendix_2.tex
\section{Derivation of the curvature operator}
\label{appendix:curvature}

An analytical expression of the magnetic equilibrium for arbitrary values of $\kappa$ and $\delta$ can be obtained by solving the Grad-Shafranov equation in the large aspect ratio limit $\epsilon=r/R_0 \rightarrow 0$, when the plasma pressure contribution is neglected \cite{Riva2017, Madden1994, Graves2013}. By keeping the zeroth and first order terms in $\epsilon$, the magnetic equilibrium takes the following form \cite{Graves2013}
\begin{eqnarray}
    \fl R_c(r, \theta) &= R_0 \Bigg[ 1+ \epsilon \cos\theta +  \sum_{m=2}^{3}\frac{S_m(r)}{R_0}\cos [(m-1)\theta]-\frac{1-m}{2\epsilon}\Bigg(\frac{S_m(r)}{R_0} \Bigg)^2\cos \theta \Bigg],\label{Equil_R} \\
    \fl Z_c(r, \theta) &= R_0 \bigg[ \epsilon \sin \theta - \sum_{m=2}^{3} \frac{S_m(r)}{R_0} \sin [(m-1)\theta] -\frac{1-m}{2\epsilon}\Bigg(\frac{S_m(r)}{R_0} \Bigg)^2\sin \theta\bigg],\label{Equil_Z}
\end{eqnarray}
where the functions $S_2(r), S_3(r)$ are related to the shaping parameters
\begin{equation}
    \kappa = \frac{a-S_2(a)}{a+S_2(a)} \qquad 
\end{equation}
and
\begin{equation}
\delta=\frac{4S_3(a)}{a}.
\end{equation}
By assuming a strongly localized RBM at $\theta=0$, it is possible to approximate the curvature operator as $\mathcal{C}^y\simeq-\partial_r R_c(r,\theta)\bigr|_{{\theta=0}}=-\partial_r R_c(r,0)-\partial_\theta R_c(r,0)\partial_r \theta\bigr|_{{\theta_*=0}}$. Then, using the fact that $\partial_r\theta\bigr|_{{\theta_*=0}}=0$ for $\epsilon=0$, the curvature coefficient at the LCFS, $r=a$, can be recast as
\begin{equation}
    \fl\mathcal{C}(\kappa, \delta, q)=\frac{\partial R_c(r,\theta)}{\partial r}\Biggr|_{{\theta_*=0}} = 1+ \sum_{m=2}^3 S'_m(a) - \sum_{m=2}^3 \frac{1-m}{a}\Bigg[S_m(a)S'_m(a)-\frac{S_m(a)^2}{2a} \Bigg].
    \label{append:curv3}
\end{equation}
The expressions for $S_m(r)$ and $q(r)$ are given by
\begin{eqnarray}
    S_m(r) &= S_m(a)\Big(\frac{r}{a}\Big)^{m-1}\frac{q(r)s(r)+2q_0\frac{m+1}{m-1}}{qs+2q_0\frac{m+1}{m-1}}, \\
    q(r) &=q_0 + (q-q_0)\bigg(\frac{r}{a}\bigg)^2,
\end{eqnarray}
where $q_0$ is the safety factor measured on the magnetic axis. 
For $m=2,3$ and $r=a$, the shaping term and its derivatives can be written as
\begin{eqnarray}
    \frac{S_2(a)}{a}&=-\frac{\kappa-1}{\kappa+1}, \\
    \frac{S_3(a)}{a}&=\frac{\delta}{4} 
\end{eqnarray}
and
\begin{eqnarray}
    S'_2(a) &= -\bigg(\frac{\kappa-1}{\kappa+1}\bigg)\bigg(\frac{3q}{q+2}\bigg), \\
    S'_3(a) &= \frac{\delta q}{q+1}.
\end{eqnarray}
Finally, by inserting the expressions of $S_m(r)$ and $S'_m(r)$ for $m=2,3$ into Eq. (\ref{append:curv3}), the curvature coefficient at $\theta=0$ leads to
\begin{equation}
    \fl\mathcal{C}(\kappa, \delta, q) =  1-\frac{\kappa-1}{\kappa+1}\frac{3q}{q+2} + \frac{\delta q}{1+q} + \frac{(\kappa-1)^2(5q-2)}{2(\kappa+1)^2(q+2)} + \frac{\delta^2}{16}\frac{7q-1}{1+q}.
\end{equation}